\documentclass{jfm}
\usepackage{amssymb,MnSymbol,mathrsfs}
\usepackage{natbib}
\usepackage{amsfonts,amsbsy}
\usepackage{graphicx}
\usepackage{color} 
\usepackage{natbib} 
\usepackage{tikz}
\usetikzlibrary{intersections,shapes.arrows}

\usepackage{lineno}

\newcommand*\patchAmsMathEnvironmentForLineno[1]{%
  \expandafter\let\csname old#1\expandafter\endcsname\csname #1\endcsname
  \expandafter\let\csname oldend#1\expandafter\endcsname\csname end#1\endcsname
  \renewenvironment{#1}%
     {\linenomath\csname old#1\endcsname}%
     {\csname oldend#1\endcsname\endlinenomath}}%
\newcommand*\patchBothAmsMathEnvironmentsForLineno[1]{%
  \patchAmsMathEnvironmentForLineno{#1}%
  \patchAmsMathEnvironmentForLineno{#1*}}%
\AtBeginDocument{%
\patchBothAmsMathEnvironmentsForLineno{equation}%
\patchBothAmsMathEnvironmentsForLineno{align}%
\patchBothAmsMathEnvironmentsForLineno{flalign}%
\patchBothAmsMathEnvironmentsForLineno{alignat}%
\patchBothAmsMathEnvironmentsForLineno{gather}%
\patchBothAmsMathEnvironmentsForLineno{multline}%
}

\DeclareFontFamily{U}{cbgreek}{}
\DeclareFontShape{U}{cbgreek}{m}{n}{
        <-6>    grmn0500
        <6-7>   grmn0600
        <7-8>   grmn0700
        <8-9>   grmn0800
        <9-10>  grmn0900
        <10-12> grmn1000
        <12-17> grmn1200Bu14
        <17->   grmn1728
      }{}
\DeclareFontShape{U}{cbgreek}{bx}{n}{
        <-6>    grxn0500
        <6-7>   grxn0600
        <7-8>   grxn0700
        <8-9>   grxn0800
        <9-10>  grxn0900
        <10-12> grxn1000
        <12-17> grxn1200
        <17->   grxn1728
      }{}

\makeatletter
\newcommand{\normalorbold}{%
  \ifnum\pdf@strcmp{\math@version}{bold}=\z@ bx\else m\fi
}
\makeatother

\def\id{\mathrm{id}}
\def\d{\mathrm{d}}

\def\beq{\begin{equation}}
\def\eeq{\end{equation}}

\def\eps{\varepsilon}

\def\bx{\boldsymbol{x}}

\def\bu{\boldsymbol{u}}

\def\bxi{\boldsymbol{\xi}}

\def\L{\mathcal{L}}
\def\ip{\iota}
\def\M{M}
\def\Diff{\mathrm{Diff}(M)}
\def\SDiff{\mathrm{SDiff}(M)} 
\def\baru{\bar u}
\def\pmo{\mathsf{p}}
\def\hmo{\mathsf{h}}
\def\P{\mathsf{P}}
\def\I{\mathsf{I}}

\def\tw{\tilde{w}}

\def\lie{\mathcal{L}}
\DeclareMathOperator{\divv}{div}

\def\argmin{\mathop{\mathrm{arg\;min}}}
\def\qflat{q_{\flat}}

\newcommand{\bs}[1]{\boldsymbol{#1}}

\newcommand{\av}[1]{\langle #1 \rangle}

\newcommand{\dt}[2]{\frac{\mathrm{d} #1}{\mathrm{d} #2}}

\newcommand{\eqn}[1]{(\ref{eqn:#1})}
\newcommand{\lab}[1]{\label{eqn:#1}}
\newcommand{\inter}[1]{\quad \textrm{#1} \quad}

\newcommand{\barL}[1]{\overline{#1}^\mathrm{L}}

\definecolor{armygreen}{rgb}{0.29, 0.33, 0.13}

\newcommand\AG[1]{{\color{magenta}#1}}

\newcommand\change[1]{#1}

\def\XXint#1#2#3{{\setbox0=\hbox{$#1{#2#3}{\int}$}
\vcenter{\hbox{$#2#3$}}\kern-.5\wd0}}

\title[Geometric generalised Lagrangian mean theories]{Geometric generalised Lagrangian mean theories}
\author[A. D. Gilbert \& J. Vanneste]{Andrew D. Gilbert$^1$ and Jacques Vanneste$^2$} 
\affiliation{$^1$ Department of Mathematics, College of Engineering, Mathematics and
Physical Sciences, University of Exeter, EX4 4QF, UK \smallskip \\ $^2$ School of Mathematics and Maxwell Institute for Mathematical Sciences, \\
University of Edinburgh, Edinburgh EH9 3FD, UK}

\begin{document}

\maketitle

\begin{abstract}
Many fluctuation-driven phenomena in fluids can be analysed effectively using the generalised Lagrangian mean (GLM) theory of \citet{AnMc78a}. This finite-amplitude theory relies on particle-following averaging to incorporate the constraints imposed by the material conservation of certain quantities in inviscid regimes. Its original formulation, in terms of Cartesian coordinates, relies implicitly on an assumed Euclidean structure; as a result, it does not have a geometrically intrinsic, coordinate-free interpretation on curved manifolds, and suffers from undesirable features. 
Motivated by this, we  develop a geometric generalisation of GLM that we formulate intrinsically using coordinate-free notation. One benefit is that the theory applies to arbitrary Riemannian manifolds; another is that it establishes a clear distinction  between results that stem directly from  geometric consistency and those that depend on particular choices. 

Starting from a decomposition of an ensemble of flow maps into mean and perturbation, we define the Lagrangian-mean momentum as the average of the pull-back of the momentum one-form by the perturbation flow maps. We show that it obeys a simple equation which guarantees the conservation of Kelvin's circulation,
irrespective of the specific definition of the mean flow map. The Lagrangian-mean momentum is the integrand in Kelvin's circulation  and distinct from the mean velocity  (the time derivative of the mean flow map) which advects the contour of integration. 
A pseudomomentum consistent with that in GLM can then be defined by subtracting the Lagrangian-mean momentum from the one-form obtained from the mean velocity   using the manifold's metric.

The definition of the mean \change{flow map} is based on choices made for reasons of convenience or aesthetics. We discuss four possible definitions: a direct extension of standard GLM, a definition based on optimal transportation, a definition based on a geodesic distance in the group of volume-preserving diffeomorphisms, and the glm definition proposed by \citet{SoRo10}. Assuming small-amplitude perturbations, we carry out order-by-order calculations to obtain explicit expressions for the mean velocity and Lagrangian-mean momentum at leading order. We also show how the wave-action conservation of GLM extends to the geometric setting.
 
To make the paper self-contained, we introduce in some detail the tools of differential geometry and main ideas of geometric fluid dynamics on which we rely. These include variational formulations which we  use for alternative derivations of some key results. We mostly focus on the Euler equations for incompressible inviscid fluids but sketch out extensions to the \change{rotating--stratified Boussinesq}, compressible Euler, and magnetohydrodynamic equations.
We illustrate our results with an application to the interaction of inertia-gravity waves with balanced mean flows in rotating--stratified fluids.

\end{abstract}

\section{Introduction}

The interactions between waves, or more broadly fluctuations, and mean flows are at the heart of a broad range of fluid-dynamical phenomena, from acoustic streaming to the formation of jets in geo- and astrophysical fluids. These phenomena involve the nonlinear rectification of fluctuations, leading to the emergence of persistent mean-flow structures. It has long been realised that they are best described using some form of Lagrangian representation of fluid motion, especially when dissipative processes play a secondary role. In essence, this is because persistent flows can often be identified with changes in vorticity which, in turn, is controlled by material transport in the absence of dissipation.  
The theoretical articulation of this observation, pioneered by \citet{De70}, \citet{Br71}, \citet{So72}, \citet{Gr75} and others, culminated in the development by \citet{AnMc78a} of the Generalised Lagrangian Mean (GLM) theory which has since become an organising centre for research on wave--mean flow interactions. We refer the reader to the book by \citet{Bu14} for a detailed exposition of the GLM formalism and of some of its more recent applications.  

One of the advantages of GLM is its generality: it considers an arbitrary ensemble of flows and an arbitrary averaging operator, and deduces from the Navier--Stokes equations or their generalisations averaged equations governing the fully nonlinear dynamics of a suitably defined Lagrangian mean flow. These can be specialised to specific settings, typically involving assumptions of multiple time scales, small amplitudes and spatial scalings, but in their generality they exhibit fundamental properties such as an averaged form of Kelvin's circulation theorem. They also capture, at once, multiple asymptotic regimes, something that standard asymptotic expansion methods do not achieve (compare, for instance, the GLM approach of \citet{BuMc98} and \citet{HoBuFe2011} with the multiscale approach of \citet{WaYo15}).

The GLM formalism has however one important deficiency, especially for incompressible fluids,
which are our main focus. Because it has been developed in Cartesian coordinates and makes
use of Euclidean parallel transport in its foundations, GLM is
constructed outside the natural configuration space of incompressible fluid dynamics, namely the space $\SDiff$ of
all volume-preserving diffeomorphisms from the fluid domain manifold $M$  to
itself.  In this sense GLM does not respect the intrinsic geometry
of fluid dynamics -- neither the finite-dimensional geometry of
$M$, nor the infinite-dimensional geometry of $\SDiff$.

This  has two consequences.  First, GLM does not apply straightforwardly
to arbitrary domain manifolds $M$.  This is not only an academic
matter: non-trivial manifolds such as the sphere or
spherical shells arise in many applications.  Attempts to use GLM
for these, using either the Euclidean structure of the embedding
space or treating local coordinates as if they were Cartesian,
lead to unsatisfactory results, with mean velocity fields that fail to
be tangent to $M$ or that depend on the choice of coordinates
(McIntyre 1980; B\"uhler 2014, \S10.2.1).

Second, even in simple Euclidean geometry, the GLM Lagrangian-mean velocity in an incompressible fluid is generically divergent, even though it is an average of divergence-free velocity fields. This phenomenon, which  has been called the
`divergence effect' (McIntyre 1988; B\"uhler 2014, \S10.2.5), implies that the GLM Lagrangian-mean velocity does not belong to the tangent space of  $\SDiff$.

To understand this state of affairs, let us recall how the Lagrangian mean flow is defined. Consider a fluid particle that is mapped by an ensemble of flows to the positions $\bx'=\bx'^\alpha$, where $\alpha$ indexes the ensemble member. (In discussing  \citeauthor{AnMc78a}'s GLM, we will follow closely their notation, in particular in omitting the label $\alpha$, thus making implicit the dependence on ensemble member.) The Lagrangian mean position $\bx$ of the particle and the (ensemble-dependent) displacement $\bxi$ are then defined by
\beq 
\change{\bx' = \bx + \bxi({\bx},t)  \inter{and}  \av{\bxi({\bx},t)} = 0,}
\lab{disGLM}
\eeq
where the brackets denote an ensemble average. The difficulty in interpreting \eqn{disGLM} on non-Euclidean manifolds is clear: particle positions are then points, not vectors, and linear operations including the addition and averaging in \eqn{disGLM} do not apply to them. The definition 
\beq
\change{\barL{\bu}(\bx,t)=\av{\bu(\bx+\bxi(\bx,t),t)}}
\lab{uLGLM}
\eeq
of the Lagrangian-mean velocity $\barL{\bu}$ is similarly problematic: the vectors $\bu(\bx+\bxi(\bx,t),t)$ on the right-hand side belong to different tangent planes for different ensemble members $\alpha$ and hence cannot be averaged in a geometrically intrinsic way. \change{As noted above, a} consequence is that $\barL{\bu}$ is not tangent to non-Euclidean manifolds even though the vector fields $\bu$ are, and another is that $\nabla \cdot \barL{\bu}$ can be non-zero even though $\nabla \cdot \bu$ is zero for each ensemble member. \change{Whether these consequences are a price worth paying for the simplicity of \eqn{disGLM}--\eqn{uLGLM} and of the resulting averaged equations of motion is a matter of opinion that will depend on the application considered.}

The present paper aims at overcoming the deficiency of GLM associated with its \change{lack of geometrically intrinsic meaning}. To do so, we develop a generalisation that rests on solid geometric foundations. The main benefits are that the theory developed (i) applies to any real (Riemannian)  manifold, (ii) is independent of the choice of coordinate system, (iii) leads to a generalised mean velocity that is divergence-free for incompressible fluids; and (iv) provides clear interpretations for the computations leading to averaged equations, distinguishing steps imposed by geometric consistency from those requiring choices to be made.

The motivations for our paper and some of our results are similar to those of \citet{RoSo06a, RoSo06b} and \change{\citet{SoRo10,SoRo14}}, building on the pioneering work of \citet{So72} \change{and on more recent developments by \citet{GjHo96} and \citet{Ho02a,Ho02b}}. One difference is that these authors restrict their attention to Euclidean space.
A more significant difference is that we formulate our theory in the language of modern coordinate-free differential geometry rather than classical tensor calculus. The advantages of a coordinate-free treatment are well established in many branches of physics \citep[e.g.][]{Fr97}: it enables compact derivations, guarantees the coordinate independence of the results, and often offers clear geometric interpretations. A drawback in the present context is that it may not be familiar to the majority of fluid dynamicists. To remedy this, the paper starts in \S\,\ref{sec:background} with a self-contained introduction to the geometric formulation of fluid dynamics, including the necessary differential geometry. This formulation, initiated by \citet{Ar66}, interprets  the Euler equations for incompressible perfect fluid motion as geodesic equations in $\SDiff$ and defines the geometric structure of incompressible fluid dynamics. Our approach is to work as much as possible within this geometric structure or later its counterpart for compressible fluids.

One of the insights of \citet{RoSo06a, RoSo06b}, reflected in our work, is that the form of the averaged equations governing the Lagrangian-mean dynamics is to a large extent independent of the specific definition chosen for the mean flow (which, in standard GLM, is set uniquely by the condition $\change{\av{\bxi}}=0$ in \eqn{disGLM}). 
A key idea here is that the mean flow is defined at the level of the flow maps -- the diffeomorphisms that send fluid positions from their initial to their current configurations --  rather than velocity fields. Thus, given an ensemble of flow maps $\phi^\alpha$ associated with perturbed fluid configurations, a single $\alpha$-independent mean flow map $\bar \phi$ is chosen to represent the ensemble. Thinking of the  $\phi^\alpha$ as forming a cluster of points in $\SDiff$, as is the case for wave-like motion, this only requires $\bar \phi$ to be located well within the cluster, in a sense that specific definitions of the mean flow map make precise. We emphasise that $\bar \phi$ is in general not constructed by application of an averaging operator since the perturbed maps $\phi^\alpha$ are nonlinear objects that cannot be averaged in
any straightforward sense. The mean velocity field associated with the mean flow map $\bar \phi$ is  naturally defined as its time derivative and is generally not a straightforward average of the perturbed velocities.

We start our analysis in \S\,\ref{sec:LMD} with a derivation of an averaged momentum equation for an arbitrary choice of the mean flow.
The derivation shows that the key element of the averaging procedure, equivalent to that of \citet{So72} and \citet{AnMc78a}, is the `pull-back'  of the dynamical equations from the perturbed 
configuration to the mean configuration. The pull-back operation, to be defined in \S\,\ref{sec:diffgeo}, brings various
fluctuating quantities into a single vector space, which allows an averaging operator to be applied to construct mean
quantities, including the Lagrangian-mean momentum one-form (or covariant vector). The Lagrangian-mean momentum appears as the integrand in the averaged Kelvin circulation theorem that follows and generalises the GLM quantity $\barL{\bu} - \textbf{\textsf{p}}$, with $\textbf{\textsf{p}}$ the GLM pseudomomentum. A more general notion of pseudomomentum emerges: it is interpreted as a one-form  that encapsulates the averaged effect of the fluctuations on the Kelvin circulation, and
measures the lack of commutation between averaging on the one
hand, and the transformation of the  mean velocity vector into a one-form using the  manifold's metric, on the other.

We consider the definition of the \change{mean flow} in \S\,\ref{sec:meandef} where we discuss four possible choices.  The first extends standard GLM
to arbitrary manifolds $M$ without rectifying the problem
of divergent mean velocity.
The other three do rectify that problem, the second using ideas
from optimal transportation, and the third -- in our view the
most natural -- relying on a geodesic distance between
flow maps in SDiff($M$) to define the mean flow map as a Riemannian centre of mass in $\SDiff$.  The fourth is the `glm'  definition recently proposed by \change{\citet{SoRo10,SoRo14}} which extends work by \citet{Ho02a,Ho02b}. This definition relates the perturbation flow maps to vector fields and  imposes a zero-average condition on the latter.

Although our theoretical development holds formally for
finite-amplitude perturbations, practical applications rest on
perturbative expansions in powers of a small amplitude parameter,
which are often truncated to the first non-trivial order.  In
\S\,\ref{sec:smallamp}, we derive coordinate expressions for such leading-order
expansions for both mean velocity and Lagrangian-mean momentum for each of the four mean-flow definitions.  We
illustrate the use of these expressions in \S\,\ref{sec:Bouss} with an
application to the effect of waves on the balanced dynamics of
a rapidly rotating, strongly stratified Boussinesq fluid, giving some insight into the manner in which the divergence effect is resolved.
 While we do not pursue perturbative computations beyond the first non-trivial order, we note that our formulation is well suited to the systematic computation of higher-order corrections.

In GLM, the equations governing the mean dynamics need to be complemented by a model for the dynamics of the fluctuations. An exact model is of course provided by subtracting the mean equations from the equations describing the full dynamics. This is rarely useful, however: in practice, simplifying assumptions are made on the perturbation dynamics, in order to obtain tractable, closed models of wave--mean flow interactions. Typical simplifications include linearisation, short-wave approximation \citep[see][and references therein]{Bu14} and heuristic closures \citep[e.g.][]{Ho02a,Ho02b}. Fundamental constraints on the perturbation dynamics need however to be respected, the most important of which is the conservation  of  wave action. \citet{AnMc78b} derived \change{an exact,} general GLM version of wave-action conservation \citep[see also][\S10.3]{Gr84,Bu14}. We obtain a geometric counterpart of their derivation in \S\,\ref{sec:waveaction}.

We conclude this introduction with three remarks. First, our paper focusses chiefly on the most basic fluid model, that is, the Euler equations for incompressible perfect fluids \change{enclosed by rigid boundaries}. Admittedly, this is not the most interesting model for wave--mean flow interactions, but we think it best to establish the main ideas and techniques in as simple a context as possible. We illustrate in \S\,\ref{sec:avother} the relevance of the theory developed to more complicated models by sketching how key results extend to \change{Boussinesq} and compressible perfect fluids, and to ideal magnetohydrodynamics (MHD); see also \S\,\ref{sec:Bouss}. We leave a complete treatment of these and other relevant models for future work. Second, we note that the developments of GLM have followed two parallel strands, with some authors basing their derivations on the equations of motion \citep[e.g.][]{So72,AnMc78a,AnMc78b,Bu14}, and others relying on variational formulations \citep[e.g.][]{De70,Br71,Gr84,GjHo96,Ho02a,Ho02b,Sa13}. We acknowledge the importance of variational arguments by providing alternative variational derivations of a few essential results. Finally, a word of caution about notation is in order: throughout the paper we have adopted the `light' notation of differential geometers (not using bold for vectors, for instance), reserving the usual notation of fluid dynamics for references to classical GLM expressions, as in \eqn{disGLM}--\eqn{uLGLM}. This enables a clear distinction  between objects that have related but different meanings in the old and new theories. Our notation also avoids some of the shorthands traditionally used in GLM. Here we identify explicitly ensemble-dependent quantities by a superscript $\alpha$,  and make a careful distinction between points in the  fluid domain $M$, and maps between such points, i.e.\ elements of $\SDiff$.

\section{Background} \label{sec:background}

In this section, we provide a brief, informal introduction to the geometric formulation of fluid dynamics pioneered by \cite{Ar66}. More recent  treatments can be found, for instance, in \cite{MaHu83}, \cite{ArKh98}, \change{\cite{HoMaRa98}},  \cite{Bo01}, \cite{Br03}, \cite{HoScSt09} \change{and \cite{BeFr17}}. To make our introduction as self-contained as possible, we start by reviewing the main tools of differential geometry needed; these are standard and described, for example, in the books by \cite{HaEl73}, \cite{Sc80} or \cite{Fr97}.

\subsection{Elements of differential geometry} \label{sec:diffgeo}

\change{In this subsection, we focus on fundamentals of differential geometry. We consider fluid motion on an $n$-dimensional manifold $M$ which at this point does not have to be equipped with a metric. The manifold $M$ is taken to be fixed, that is, time-independent, and to have fixed boundary $\partial M$ (which could be empty). Free-surface effects are therefore ignored, though they could be incorporated in essentially the same formalism.  Obvious examples of manifolds $M$ of interest are the Euclidean space $\mathbb{R}^n$, with $n=2,3$, the torus $\mathbb{T}^n$, the
2-sphere $\mathbb{S}^2 \subset \mathbb{R}^3$ given by  $r=1$ and the spherical shells $r_0\leq r\leq r_1$ in $\mathbb{R}^3$, where $r$ is the radial component of spherical polar coordinates. We treat $M$ in a coordinate-independent way but, since we deal with non-relativistic fluids, retain the special role of the time coordinate $t$.}

The basic object in fluid mechanics, and more generally in continuum mechanics, is the flow map, $\phi_t$, which maps the positions of fluid particles  at the initial time $t=0$ to their positions at time $t$. Thus, if $x$ denotes the position at time $t$ of the particle initially at $a$, 
\beq
x =\phi_t a.
\eeq 
The flow map is a diffeomorphism, that is, a smooth invertible map from \change{$M$ to itself}. We write $\phi_t :  M \to M$ and $\phi_t \in \Diff$, where $\Diff$ is the set of all diffeomorphisms from $M$ to $M$. We shall be mostly be concerned with the subset $\SDiff$ of $\Diff$ consisting of all the \textit{volume-preserving} diffeomorphisms from $M$ to $M$.
For convenience, we generally omit the subscript $t$ from the flow map in what follows, simply remembering that $\phi$ depends on time, as essentially do most of the objects introduced. 
Being fixed, the boundary $\partial M$ is preserved under the action of any flow map $\phi$; if it has more than one connected component, $\phi$ must map each component to itself; otherwise, $\phi$ would be not be continuously connected to the identity map, the flow map at $t=0$.

\begin{figure}
\begin{center}

\begin{tikzpicture}[scale=0.9]

\shade[left color=gray!10,right color=gray!90] 
  (1.5,1) to[out=-10,in=100] (6,-2.0) -- (9.55,1) to[out=100,in=-10] (5.45,4) -- cycle;

\coordinate (A) at (6.1,1.6);
\coordinate (B) at (5,3);
\coordinate (C) at (9,1.0);
\coordinate (D) at (6.1,2.2);

\draw [solid,thick] (A) to[out=10,in=140] (C);

\draw [fill=gray!80]  (C) circle [radius=0.06];


\draw [fill=yellow,opacity=0.5] (3.8,1.5)--(6,-.5)--(8.5,1.5)--(6.3,3.5) --cycle;
\draw [fill=gray!80]  (A) circle [radius=0.06];
\node [below] at (A) {$x$};


\draw [thick,>=latex,->,blue] (A)-- node[above] {$v(x)$} ++(10:1.7);

\node[] at (6.3,2.8) {$T_x M$};
\node at (2.6,1.3) {$M$};
\node [left] at (C) {$\psi_s x$};

\end{tikzpicture}

\caption{
\change{A vector $v(x)$ in the tangent space $T_xM$ at $x\in M$ is defined by a curve $\psi_s x$ according to \eqn{vpsis}.}
} \label{fig:vectorinM}
\end{center}
\end{figure}
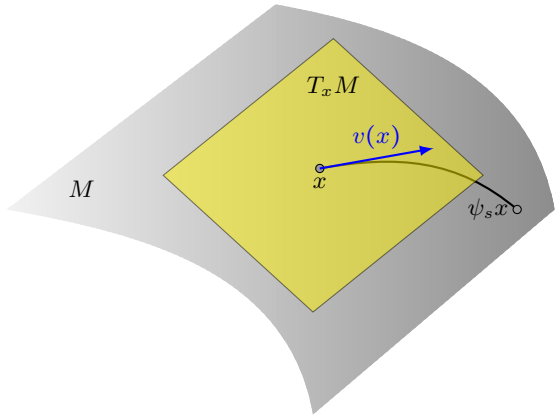

In the standard Eulerian description of fluids, $\phi$ does not appear explicitly; instead, it is the velocity field \change{$u=u(\cdot)$ that appears as the fundamental dynamical variable. It is defined by
\beq
\dot \phi a = u(x) = u(\phi a),
\lab{velocity}
\eeq
where the overdot denotes time derivative and $a \in M$ is arbitrary. Equivalently,
\beq
u = \dot \phi \circ \phi^{-1},
\lab{velocity1}
\eeq}
where $\circ$ denotes composition.
In writing \eqn{velocity} we have adhered to our convention of omitting to make explicit the time dependence of $u$ on $t$.
When the velocity field is known, the flow map is deduced by integrating the ordinary differential equations \eqn{velocity} with the initial condition $\left. \phi \right|_{t=0} = \id$, \change{that is, the identity map}. 

\change{The velocity field is an example of a vector field on $M$. In general, a vector at $x$ such as $v(x)$  can be identified as an equivalence class of curves through $x \in M$ sharing the same tangent. If $\psi_s x$ is one such curve parameterised by $s$ and with $\psi_0 x =x$, we  write that
\beq
v(x) =  \left.  \frac{\d}{\d s} \right|_{s=0} \! \psi_s x \, \in T_x M,
\lab{vpsis}
\eeq
where $T_x M$ is the tangent space to $M$ at $x$ (the set of all vectors at $x$). This formalises the interpretation of vectors as displacements -- arrows -- between two infinitesimally close points $x$ and $y=\psi_s x$ with $s \to 0$, as  illustrated in figure \ref{fig:vectorinM}.
(Note that \eqn{velocity} is an instance of \eqn{vpsis} with $\psi_s = \phi_{t+s}$.)
 A vector field, then, is a map \change{$v$} from $M$ to the tangent bundle $T M$ (the union of the tangent spaces $T_x M$ at all points $x\in M$) with the property that $v(x) \in T_x M$.
We emphasise this property because traditional GLM formulations involve objects such as $u \circ \phi$ which are not vector fields (since $(u \circ \phi)(x) = u(\phi x) \in T_{\phi x} M\not= T_x M$) and, in our opinion, are best avoided for clarity in a general setting.}
Note that if $M$ has a fixed boundary $\partial M$, the velocity field is tangential to it, $u \parallel \partial M$ (meaning that $u(x) \in T_x \partial M$ for boundary points $x \in \partial M$).

Fluid dynamics on  arbitrary manifolds is formulated most naturally using fields of differential one-forms in addition to vector fields. A one-form at $x$ is defined as a linear map from  vectors in $T_xM$ to the real numbers. One-forms at $x$ make up a vector space, the cotangent space $T^*_xM$, and their application to vectors of $T_xM$ is interpreted as a pairing (or interior product) which produces a real number. 
If $v$ is a vector field on $M$ and $\nu$ is a field of one-forms, the pairing at $x$ is denoted as
\beq
\nu(x)(v(x))= \nu(v)(x) \in \mathbb{R}
\lab{ip}
\eeq
and defines a scalar field, a real function  over the whole manifold $M$. In coordinates, with $v^i(x)$ and $\nu_i(x)$ denoting the components of the vector and one-form, the pairing \eqn{ip} is simply given by $\nu(v) = v^i \nu_i$ at each point $x$. Note that the coordinates $x^i$ need not be orthogonal; indeed `orthogonality' is meaningless since no metric has yet been introduced. Pictorially, one-forms can be represented as stacks of infinitesimally close planes (in dimension $n=3$ -- hyperplanes of dimension $n-1$ in general) and their pairing with a vector interpreted as measuring the number of planes pierced by the vector.
(Note that, although we will not adopt this terminology, vectors such as $v(x)$ are traditionally referred to as `contravariant vectors' and one-forms such as $\nu(x)$ as `covariant vectors', and their pairing as `contraction'.)

Higher-order forms (and fields thereof) are defined similarly as linear maps on sets of vectors, with added skew-symmetry properties\change{: a $k$-form $\alpha$ acts linearly on $k$ vectors $v_1,\cdots,v_k$ to produce the real number $\alpha(v_1,\cdots,v_k)$ which is invariant under even permutations of its arguments and flips signs under odd permutations. For example, a two-form $\alpha$ acts linearly on pairs of vectors $u$, $v$, with $\alpha(u,v)=-\alpha(v,u) \in \mathbb{R}$; in coordinates, it has components $\alpha_{ij}$ that are anti-symmetric, $\alpha_{ij} = - \alpha_{ji}$, and satisfy $\alpha(u,v)=\alpha_{ij} u^i v^j$. Zero-forms are simply scalars. For an $n$-dimensional $M$, the set of $k$-forms makes up a vector space of dimension $n \choose k$, so that all $n$-forms are scalar multiples of one another and therefore proportional to a reference $n$-form, termed the `volume-form' and denoted by $\omega$. Differential forms may be naturally integrated, along curves for one-forms, surfaces for two-forms, etc.\ \cite[e.g.][]{Sc80,Fr97}.} The wedge (or exterior) product $\wedge$ combines a $k$-form $\alpha$ with an $l$-form $\beta$ to produce a $(k+l)$-form  $\alpha\wedge\beta$ \change{and is defined as an anti-symmetrised tensor product}.

\begin{figure}
\begin{center}

\begin{tikzpicture}[scale=0.9]

\fill [ball color=white]
(0,0) .. controls +(0,2)  and +(-1,0)  .. (4,4) .. controls +(1,0) and +(0,2) .. 
(7,1.4) .. controls +(0,-2) and +(0,-2) .. (0,0);

\coordinate (A) at (1.5,0.5);
\coordinate (B) at (5,3);
\coordinate (C) at (2.9,0.3);
\coordinate (D) at (6.1,2.2);

\draw [dashed,thick] (A) to[out=-20,in=180] (C);

\draw [dashed,thick] (B) to[out=-5,in=120] (D);

\draw [fill=yellow,opacity=0.5] (4.2,3.1)--(5.5,3.2)--(6.7,2.6)--(5.5,2.5)--cycle;
\draw [fill=yellow,opacity=0.5] (1,0)--(1.6,1.2)--(3.4,0.4)--(2.8,-0.8) --cycle;
\draw [fill=gray!80]  (A) circle [radius=0.06];
\node [below] at (A) {$x$};
\draw [fill=red]  (B) circle [radius=0.06];
\node [right] at (C) {$y$};
\draw [fill=red]  (D) circle [radius=0.06];
\draw [fill=gray!80]  (C) circle [radius=0.06];

\draw[>=stealth,shorten >=2pt,->,thick] (A) to[out=50,in=150] node[above] {$\phi$} (B) ;
\draw[>=stealth,shorten >=2pt,->,thick] (C) to[out=50,in=180] node[above] {$\phi$} (D) ;

\draw [thick,>=latex,->,blue] (A)-- node[below] {$v(x)$} ++(-20:1.5);
\draw [thick,>=latex,->,blue] (B)-- node[above] {$\phi_* v(x)$} ++(-10:1.2);

\node[above] at (4.5,2.2) {$T_{\phi x} M$};
\node[] at (1.3,-0.5) {$T_x M$};
\node at (6,1) {$M$};


\end{tikzpicture}

\vspace{-3ex}
\caption{Push-forward of a vector $v(x)$ at $x$: a curve $\psi_s x$
tangent to $v(x)$ at $x$ (dashed line) which includes a point $y$ is mapped to a curve passing through $\phi x$ and $\phi y$. The tangent of the latter curve at $\phi x$ defines the push-forward $\phi_* v(x)$. 
} \label{fig:push}
\end{center}
\end{figure}
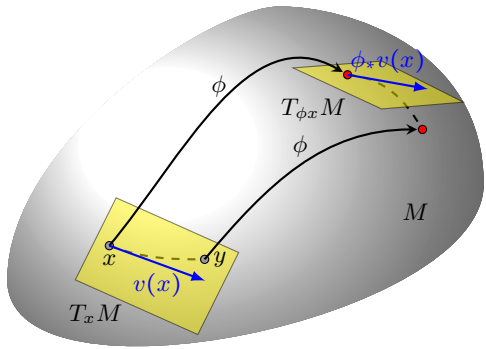

The geometric tools used in this paper include the push-forward and pull-back of vectors and differential forms (and indeed any tensor) by diffeomorphisms such as the flow map. The definition of the push-forward $\phi_*$ by a map $\phi:M\to M$ recognises that the action of diffeomorphisms on points naturally induces an action on vectors thought of as infinitesimal displacements. 
The push-forward $\phi_* v(x) \in T_{\phi x}M$  of $v(x)$ is a vector at the image point $\phi x$ defined as the tangent to \change{the curve $\phi(\psi_s x)$, with $\psi_s x$ a curve tangent to $v$ at $x$ to satisfy \eqn{vpsis}; explicitly, we have}
\beq
\phi_* v(x) = \left. \dt{}{s} \right|_{s=0} \phi(\psi_s x) ; 
\eeq 
see figure \ref{fig:push}.
The push-forward $\phi_*$, often termed the differential of $\phi$, is linear. It provides a way to map entire vector fields $v$
according to
\beq
(\phi_* v)(\phi x) = \phi_* v(x),
\eeq
and arises naturally in applications of the chain rule, for example in \eqn{key} below. 
In coordinates $x^i$, the push-forward takes the form
\beq
(\phi_* v)^i(\phi x) = v^j(x) \, \partial_{j} \phi^i(x).
\lab{pushcoord}
\eeq

A well-known example of the push-forward of a vector field arises in MHD with the observation that, \change{for incompressible flow} in the absence of diffusion, the magnetic field is a `frozen' vector field, transported by the flow \change{according to the Cauchy solution}. The magnetic field $b$ at any $t$ is then the push-forward of the initial magnetic field $b_0$ by the \change{volume-preserving} flow map: $b=\phi_* b_0$. 

Once the push-forward of a vector is defined, the definitions of the push-forward and pull-back of various tensors follow naturally. The pull-back $\phi^*$ of a vector \change{$v(x) \in T_x M$} by $\phi$ is simply the push-forward under the inverse map $\phi^{-1}$, that is,
\begin{equation}
\change{\phi^* v(x) = (\phi^{-1})_* v(x)} \in T_{\phi^{-1} x}M, 
\lab{pullbackvec}
\end{equation}
using the invertibility of the diffeomorphism. The pull-back of one-forms is defined using their duality with vectors: if $\change{\nu(x)} \in T^*_xM$ is a one-form, its pull-back $\change{\phi^* \nu(x)} \in T^*_{\phi^{-1} x}M$ satisfies
\beq
\change{(\phi^* \nu)(v)=\nu(\phi_* v)} \ \ \textrm{for any} \ \ v \in T_{\phi^{-1} x}M.
\eeq
With this definition, the coordinate expression of the pull-back of a one-form field $\nu = \nu_i \, \d x^i$ is readily obtained from \eqn{pushcoord} as
\begin{equation}
(\phi^* \nu)_i (\phi^{-1}x) = \nu_j(x) \, \partial_i \phi^j(\phi^{-1}x).
\lab{pullbackform}
\end{equation}
The pull-back of other differential forms is defined similarly via pairing with vectors; the push-forward is defined as the pull-back by the inverse map in a way that mirrors \eqn{pullbackvec}.   
Note that the pull-back and push-forward of scalar functions (zero-forms) are simple compositions: $\phi^* f = f \circ \phi$ and $\phi_* f = f \circ \phi^{-1}$.

Another important differential-geometric tool is the Lie derivative with respect to a vector field $v$.  For any tensor field (vector or form) $\tau$, the Lie derivative is defined by
\beq
\lie_v \tau = \left. \dt{}{s} \right|_{s=0} \psi_s^* \tau,
\lab{lie}
\eeq
where the map $\psi_s$ is the flow of $v$ in the sense that 
\beq
\dt{}{s} \psi_s x= v(\psi_s x), \quad \psi_0 = \id.
\eeq
An application to the fluid flow map $\phi$ and associated (time-dependent) vector field $u$ gives the useful formula
\beq
\partial_t (\phi^* \tau) = \phi^* \left( \partial_t \tau + \lie_u \tau \right), 
\lab{dtpullback}
\eeq 
holding when $\tau$ depends explicitly on $t$. Another useful property concerns the action of pull-back on Lie derivatives: for any vector field $v$, 
\beq
\phi^* (\lie_v \tau) = \lie_{\phi^* v} (\phi^* \tau).
\lab{pullbacklie}
\eeq
Both properties \eqn{dtpullback} and \eqn{pullbacklie} are \change{derived} in appendix \ref{sec:pullback}, \change{and in \citet{SoRo10,SoRo14} using coordinate methods.}
Coordinate expressions for Lie derivatives are obtained by writing the map $\psi_s$ in \eqn{lie} as 
$(\psi_s x)^i=x^i + s v^i(x) + o(s)$
 and using the coordinate expression of the pull-back. This gives
\beq \lab{liecoord}
\change{
\lie_v f = v^j \partial_j f , }
\quad \quad
(\lie_v u)^i = v^j \partial_j u^i  - u^j \partial_j v^i , 
\quad \quad 
(\lie_v \nu)_i =  v^j \partial_j \nu_i + \nu_j \partial_i v^j
\eeq
for the Lie derivatives of scalar fields $f$ (zero-forms), vector fields $v$ and one-form fields $\nu$, respectively. \change{An interpretation of the Lie derivative of a vector field is used repeatedly in what follows: if a map $\psi_{s,t}$ depends on the two parameters $s$ and $t$, we can define the two $s$- and $t$-dependent vector fields
\beq
u=\dt{}{t} \psi_{s,t} \circ \psi_{s,t}^{-1} \inter{and} v = \dt{}{s} \psi_{s,t} \circ \psi_{s,t}^{-1},
\eeq
in the manner of \eqn{velocity}. The Lie derivative of $v$ with respect to $u$ then satisfies
\beq \lab{liecommute}
\lie_u v = - \lie_v u = \partial_s u - \partial_t v.
\eeq
}

The final operation needed in what follows is the exterior derivative $\d $ which acts on a $k$-form \change{field} to produce a $(k+1)$-form \change{field}. It is defined by its \change{action on a scalar field $f$, giving a one-form field $\d f$ with the property that $(\d f)(v) =\lie_v f $ for any vector field $v$},
by the property that \change{$\d^2\alpha = 0$ for any form $\alpha$}, and by its action on the exterior (wedge) products of forms, $\d(\alpha\wedge\beta) = \d\alpha\wedge\beta + (-1)^k \alpha\wedge\d\beta$ where $\alpha$ is a $k$-form. For our purpose, an important property of the exterior derivative is that it commutes with \change{pull-backs/push-forwards by diffeomorphisms  and with Lie derivatives:
\beq
\phi_* (\d \alpha)  = \d ( \phi_* \alpha), \quad
\phi^* (\d \alpha)= \d (\phi^* \alpha), \quad
\d (\lie_v \alpha) = \lie_v (\d \alpha) . 
\lab{usefulprops}
\eeq}
Another is Cartan's formula 
\beq
\lie_v \alpha = \ip_v \d \alpha + \d (\ip_v \alpha) , 
\lab{cartan}
\eeq
relating the Lie derivative of a $k$-form $\alpha$ to its exterior derivative by means of the interior product $\ip_v \alpha$ defined as the $(k-1)$-form such that $(\ip_v \alpha)(v_1,\ldots,v_{k-1}) = \alpha(v,v_1,\ldots,v_{k-1})$ for all vectors $v_1,\ldots,v_{k-1}$.

\subsection{Fluid dynamics on a \change{Riemannian} manifold}

We now review the geometric formulation of fluid dynamics which naturally emerges from \citeauthor{Ar66}'s (\citeyear{Ar66}) interpretation of the motion of a perfect incompressible fluid on an arbitrary manifold as geodesic motion \cite[see also][]{EbMa70,HoMaRa98,ArKh98}. This requires the manifold $M$ to have a Riemannian structure, that is, to be equipped with a metric $g(\cdot,\cdot)$, a positive-definite bilinear form which, at each point $x \in M$, maps pairs of vectors to non-negative real numbers, thus generalising the scalar product of Euclidean space. The metric induces a preferred $n$-form, which we take as the reference volume form $\omega$. This volume form defines the volume element on $M$ and appears naturally in integrals over the domain such as $\int_M f \omega$ where $f$ is a scalar function. A fluid motion is incompressible provided the associated flow map $\phi$ preserves volume in the sense that
\beq \lab{volume}
\phi^* \omega = \omega.
\eeq
This is the explicit form of $\phi \in \SDiff \subset \Diff$.

The metric $g$ induces a distance $d(x,y)$ between points $x$ and $y$ on $M$. This is the length of the shortest path from $x$ to $y$, which is defined by
\beq
\change{d(x,y) = \inf_{\gamma_s : [0,S] \to \M} \int_0^{S}  \left[ 
g( \gamma'_s,\gamma'_s) \right]^{1/2}  \, \d s, \quad \gamma_0=x,  \ \ \gamma_S=y,}
\lab{d2sqrt}
\eeq
where $\gamma_s$ denotes paths joining $x$ and $y$, and the \change{prime} denotes an $s$-derivative. The distance here does not depend on the parameterisation of the path, including the choice of interval $[0,S]$. It is then a standard result that obtaining $d(x,y)$ by the above variational principle is equivalent to a more convenient formula for the distance squared,
\beq
d^2(x,y) = \inf_{\gamma_s : [0,1] \to \M} \int_0^{1}   
g( \gamma'_s, \gamma'_s)  \, \d s, \quad \gamma_0=x,  \ \ \gamma_1=y. 
\lab{d2}
\eeq
Here the parameterisation is no longer arbitrary, and the minimising path $\gamma_s$ is characterised by a constant kinetic energy 
\change{$\tfrac{1}{2} g( \gamma'_s, \gamma'_s)$}. We refer the reader to \cite{ArKh98} for more discussion; we write other types of distances using fomulae of the type \eqn{d2} in what follows without further comment.

The distance $d(x,y)$ in \eqn{d2sqrt} or \eqn{d2} between points in $M$ does not play an immediate role in the dynamics of fluids on $M$. The key role is rather played by the distance between two points in $\SDiff$, that is, between pairs of volume-preserving diffeomorphisms $\phi, \, \psi$. This distance is defined by
\begin{equation} 
D^2(\phi,\psi) = \inf_{\gamma_s  : [0,1]  \to \SDiff} \int_0^1 \int_\M g( \gamma'_s,  \gamma'_s) \omega \, \d s, \quad \gamma_0 = \phi, \ \ \gamma_1 = \psi,
\lab{D2} 
\end{equation}
where $\gamma_s$ now denotes paths joining $\phi$ to $\psi$ in $\SDiff$. Introducing the $s$-dependent vector field \change{$q =  \gamma'_s \circ \gamma_s^{-1}$} and using volume preservation, this can be rewritten as 
\begin{equation}
D^2(\phi,\psi)= \inf_{\gamma_s  : [0,1]  \to \SDiff} \int_0^1 \int_\M g(q, q) \omega \, \d s, \quad \gamma_0 = \phi, \ \ \gamma_1 = \psi. \lab{D22}
\end{equation}
This makes explicit the right-invariance of the metric associated with $D^2$, that is, the invariance of the integral over $M$ in \eqn{D2} with respect to the right-translations $\gamma_s \mapsto \gamma_s \circ \chi$ for an arbitrary $s$-independent $\chi \in \SDiff$. 
This invariance implies that $D(\phi,\psi)=D(\phi \circ \psi^{-1},\mathrm{id})$.
In physical terms, $D^2$ is twice the $s$-integrated kinetic energy for a flow that joins the two configurations $\psi$ and $\phi$ of the fluid. 

The geometric view of the Euler equations for an incompressible perfect fluid on $M$ is that they arise as geodesic equations on $\SDiff$ equipped with the distance \eqn{D22}. Specifically, the Euler equations are obtained by minimising the action 
\beq
\mathscr{A}[\phi] = \tfrac{1}{2} \int_0^T  \int_\M g(u, u) \omega \, \d t
\lab{action}
\eeq
over all time-dependent flow maps $\phi \in \SDiff$ (with corresponding velocity $u = \dot \phi \circ  \phi^{-1}$) which join two fixed flow maps between $t=0$ and $t=T$. 
The motion which achieves the minimum is an incompressible flow with constant kinetic energy. 
When the Euler equations are derived in this fashion, the natural dynamical variable is not the velocity field $u = \dot \phi  \circ \phi^{-1}$ but the momentum conjugate to it, that is, the one-form $\nu$ obtained from $u$ using the metric $g$ according to
\begin{equation}
\nu = g(u,\cdot) = u_\flat, \lab{inertia}
\end{equation}
say, meaning that $\nu(v)=g(u,v)$ for all $v \in T_x M$. Equation \eqn{inertia} defines the so-called inertia operator which associates a one-form such as $\nu$ to a vector such as $u$; we say that $\nu$ is dual to $u$. In coordinates \eqn{inertia} amounts to the familiar lowering of an index: $\nu_i = g_{ij} u^j$. This observation motivates the musical notation $u_\flat$ (and its inverse $\nu_\sharp$).  The one-form \eqn{inertia} dual to the velocity field will be called the `momentum one-form'.

Minimising the action $\mathscr{A}$ in \eqn{action} leads to the Euler equations in the form 
\beq \lab{euler}
\partial_t \nu + \mathcal{L}_u \nu = - \,  \d \pi, \quad \divv u  =0 , 
\eeq
where $\pi$ is a Lagrange multiplier enforcing volume preservation and $\divv u$ is defined by $\lie_u \omega  = ( \divv u ) \omega$ (see appendix \ref{sec:varieuler} for a derivation, \change{and \cite{HoMaRa98} for a general formulation applicable to a broad class of fluid models}). 
In this derivation, $\nu$ appears as the variational derivative of the Lagrangian $\int_M g(u,u) \omega$ with respect to $u$, hence its interpretation as the conjugate momentum, noted above. 
Applying the metric, \eqn{euler} can be rewritten in terms of the vector field $u = \nu_\sharp$ as 
\beq \lab{euler2}
\partial_t u + \nabla_u u = - \nabla p, \quad \divv u  =0,
\eeq
where $p=\pi+\frac{1}{2} g(u,u)$ is the usual pressure,  \change{$\nabla p = (\d p)_\sharp$} and $\nabla_u$ is the covariant derivative based on $g$ \change{(see \eqn{eulercoordu} below for a coordinate expression)}. The derivation of \eqn{euler2} uses the identity 
\beq
\lie_u \nu = ( \nabla_u u)_\flat + \tfrac{1}{2} \d g(u,u) 
\lab{liecov}
\eeq
\citep[e.g.][Theorem IV.1.17]{ArKh98}. While \eqn{euler2} may have the more familiar form, \eqn{euler} turns out to be superior in the context of GLM theories. Its advantages are hinted at by the ease with which Kelvin's circulation theorem follows, as we now show. 
Using \eqn{dtpullback} and \eqn{usefulprops}, \eqn{euler} can be rewritten as
\beq
\partial_t ( \phi^* \nu) = - \, \d (\phi^* \pi).
\eeq
Now, let $C_0 \subset M$ be a closed curve in the initial fluid configuration and $\phi C_0$ its  image as deformed under the flow map $\phi$, then
\beq \lab{kelvin}
\dt{}{t} \oint_{\phi C_0} \nu = \dt{}{t} \oint_{C_0} \phi^* \nu= \oint_{C_0} \partial_t \left(\phi^* \nu \right)= - \oint_{C_0} \d \left(\phi^* \pi\right) = 0.
\eeq
This shows that the circulation around any material curve
\beq\lab{Kelv}
\oint_{\phi C_0} \nu = \mathrm{const}.
\eeq
The vorticity is the two-form $\d \nu$. It is transported by the flow: taking the exterior derivative $\d$ of \eqn{euler} and using that $\d^2=0$  gives the vorticity equation 
\beq\lab{vorteqn}
\partial_t \d \nu + \lie_u \d \nu =0. 
\eeq
\change{This means that vorticity is pushed forward by the flow map from the initial condition, $ \d\nu = \phi_*  \d\nu |_{t=0}$; see \cite{BeFr17} for discussion of this and applications to helicity.} 

For completeness, we record the coordinate version of \eqn{euler} and \eqn{euler2}. Using \eqn{liecoord}, \eqn{euler} is 
written explicitly as
\beq \lab{eulercoord}
\partial_t \nu_i + u^j \partial_j \nu_i + \nu_j \partial_i u^j = - \, \partial_i \pi, \quad |g|^{-1/2} \,\partial_i \left(|g|^{1/2}\, u^i \right) =0,
\eeq
where the expression for the divergence is obtained using that $\omega = |g|^{1/2} \,\d x^1 \wedge \d x^2 \wedge \cdots \wedge \d x^n$, with $|g|$ the determinant of $g_{ij}$. Introducing $\nu_i = g_{ij} u^j$ into the first equation of \eqn{eulercoord} leads after a short computation to 
\beq \lab{eulercoordu}
\partial_t u^i + u^j \left(\partial_j u^i  + \Gamma^i_{jk} u^k\right) =  - g^{ij} \partial_j p,
\eeq
where $\Gamma^i_{jk} = \frac{1}{2} g^{il} (\partial_j g_{lk} + \partial_k g_{lj} - \partial_l g_{jk})$ is the Christoffel symbol \citep{HaEl73,Sc80}. This is the coordinate version of \eqn{euler2}.

\change{\subsection{Other fluid models}} \label{sec:othermodels}

To conclude this section we note that more complicated fluids than the perfect incompressible fluids considered so far are governed by equations that have a  structure similar to that of the Euler equations. \change{The three-dimensional ($n=3$) Boussinesq equations are widely used to model rotating stratified geophysical fluids. In coordinate-free notation, they take the form 
\begin{subequations} \lab{boussinesq}
\begin{align}
\partial_t \nu_\mathrm{a} + \lie_{u} \nu_\mathrm{a} &= - \d \pi + \theta\,  \d \Phi, \lab{boussimom} \\
\partial_t \theta + \lie_{u} \theta &= 0, \lab{boussibuo} \\
\divv u &= 0, \lab{boussinc}
\end{align}
\end{subequations}
where $\theta$ is the buoyancy acceleration, $\Phi$ the geopotential divided by $g$, given by $z$ in the Euclidean case, and $\nu_\mathrm{a}$ is the absolute momentum one-form
\beq
\nu_\mathrm{a} = \nu + A.
\eeq
Here $A$ is a one-form that encodes planetary rotation; on the $f$-plane, with $f$ the constant Coriolis parameter, and in Cartesian coordinates it is given by 
\beq
A = \tfrac{1}{2}f  (x \, \d y - y \, \d x),
\eeq
up to an irrelevant differential. Like the Euler equations, the Boussinesq equations have a variational formulation:  \eqn{boussinesq} may be obtained by minimising the action
\beq
\mathscr{A}[\phi] =  \int_0^T  \int_\M  \left( \tfrac{1}{2} g(u, u) + A(u) - \theta \Phi \right)  \omega \, \d t
\lab{actionBouss}
\eeq
over time-dependent flow maps $\phi$ and with $\theta=\phi_* \theta_0$ transported from its initial value \citep[e.g.][]{HoMaRa98}.

Taking the exterior derivatives of \eqn{boussimom} and \eqn{boussibuo} gives
\beq
\left(\partial_t  + \lie_{u} \right) \d \nu_\mathrm{a} = \d \theta \wedge \d \Phi 
\inter{and}
\left(\partial_t  + \lie_{u} \right) \d \theta = 0.
\eeq 
The two-form $\d \nu_\mathrm{a}$ corresponds to the absolute vorticity. The conservation 
\beq
\left(\partial_t  + \lie_{u} \right) \mathcal{Q} = 0 \lab{boussiPVcons}
\eeq
of the potential-vorticity three-form 
\beq
\mathcal{Q} = \d \nu_\mathrm{a} \wedge \d \theta \lab{boussiPV}
\eeq follows immediately. The more familiar scalar Rossby--Ertel potential vorticity $Q$ is defined by $Q \, \omega = \mathcal{Q}$ and is also conserved: 
$\left(\partial_t  + \lie_{u} \right) Q = 0$. We note  that $\mathcal{Q}$ can be interpreted as the mass of \citeauthor{HaMc87}'s \citeyearpar{HaMc87} potential-vorticity substance and its impermeability property related to the alternative form $\mathcal{Q}= \d \left( \nu_\mathrm{a} \wedge \d \theta \right)$ \change{\citep[see also][]{HaMc90,BrSc93}}. 
}

\change{Similarly, a }perfect compressible fluid, with equation of state $p =P(\rho,\theta)$, where $\rho$ and $\theta$ are the density and entropy, is governed by the equations
\begin{subequations} \lab{compeulerouter}
\begin{align} \lab{compeuler}
\partial_t \nu + \lie_u \nu &= - \, \d p / \rho + \tfrac{1}{2} \d g(u,u), \\
\partial_t m + \lie_u m &= 0, \lab{masscomp} \\
\partial_t \theta + \lie_u \theta &= 0. 
\end{align}
\end{subequations}
Here the mass $m$ is defined as the $n$-form $m = \rho \omega$, so that \eqn{masscomp} is equivalent to the continuity equation $\partial_t \rho + \divv (\rho u) = 0$. The conservation
\beq \lab{PV}
(\partial_t + \lie_u) \d \nu \wedge \d \theta = 0
\eeq
of the potential vorticity three-form $\d \nu \wedge \d \theta$ follows from observing that $\d p \wedge \d \rho \wedge \d \theta=0$ by virtue of the dependence $p = p(\rho,\theta)$. In three dimensions, $n=3$,  the more usual scalar Rossby--Ertel potential vorticity $Q$ is defined by the relation $Q m = \d \nu \wedge \d \theta $ \citep{Sc80}.  Its conservation (as a scalar) is an immediate consequence of the conservation of the two three-forms $\d \nu \wedge \d \theta$ and $m$.  

\change{Finally, we formulate the equations governing the dynamics of ideal MHD flows in a coordinate-free way, in $n=3$ dimensions \citep[e.g.][]{Fr97}. We let $b$ be the magnetic (vector) field, $\beta = g(b, \cdot) = b_\flat$ its dual one-form, and the flux two-form $B$ be defined by the interior product $\iota_b \omega = \omega(b,\cdot,\cdot) = B$.  
The condition that the magnetic field be divergence-free, $\divv b = 0$, amounts to requiring that $\d B = 0$  (the two-form $B$ is said to be closed). This is preserved under the induction equation, which is simply transport of $B$ by the flow $u$, 
\begin{equation}
\partial_t B + \lie_u B = 0. 
\lab{MHD1}
\end{equation}
The corresponding equation for magnetic field is then 
\begin{equation}
\partial_t b + \lie_u b + (\divv u) b = 0 . 
\lab{MHD1a}
\end{equation}
This extra term ensures that $b$ remains divergence free and means that $b$ is transported as a `tensor of weight $-1$',  as is density $\rho$ 
(see below \eqn{compeulerouter}). We will not use the machinery of tensor weights in this paper; see the parallel development in \cite{SoRo14}
and note that if the flow is incompressible, $\divv u = 0 $ and $b$ is transported as a vector field. 

The momentum equation may be written, in general, as 
\begin{align}
\rho(\partial_t \nu + \lie_u \nu) &=  \iota_b \, \d \beta  - \d p  +  \tfrac{1}{2} \rho\, \d g(u,u)  \notag \\
& = \lie_b  \beta - \d g (b,b) - \d p   +  \tfrac{1}{2} \rho \,\d g(u,u) . 
\lab{MHD2}
\end{align}
The two versions of the Lorentz force term are linked by the Cartan formula, giving here
$\lie_b \beta = \iota_b \d \beta + \d (\iota_b \beta)$. Compressible MHD is governed then by \eqn{MHD1} or \eqn{MHD1a}, \eqn{MHD2}, conservation of mass \eqn{masscomp} and an equation of state such as $p = P(\rho)$. Note that the Lorentz force term involves both the magnetic (vector) field $b$ and the corresponding one-form $\beta$ as discussed by \cite{Ho02b}, \cite{RoSo06a}.}

\section{Lagrangian-mean dynamics} \label{sec:LMD}

\subsection{Generalising GLM to arbitrary manifolds}

GLM and other Lagrangian-mean theories can be thought of as extensions to $\SDiff$ of the classical (Krylov--Bogoliubov--Mitropolsky) averaging of ODEs \citep[e.g.][]{Na73,SaVe85}. \change{The central idea is to seek a variable transformation that represents the state of the system} as the sum of a mean position and a perturbation. Crucially the mean is defined, if possible, in such a way that the perturbation remains bounded and small for long times, so that the fluctuating states stay close to the mean state. The mean dynamics is of course affected by the perturbation; as a result, the equations governing the mean dynamics contain terms representing the average effect of the perturbation. Often these terms depend only on a few features of the perturbation so there is no need to work out all the details of its evolution. 

In a similar manner, in GLM one wishes to define a mean configuration of the fluid -- a mean flow map $\bar \phi$ -- that stays close to the exact fluctuating flow maps but evolves \change{in a simpler manner} because it filters out \change{fluctuations}. \change{The theory applies to different types of averaging: the fast-time averaging that appears in asymptotic treatments of two-time-scale models, averaging along a particular direction, or a probabilistic average when fluctuations model  uncertainty}. For this reason it is customary since \citet{AnMc78a} to develop GLM theories for an arbitrary ensemble average, only stipulating the ensemble when applications are made specific. We follow this approach and consider an ensemble of flow maps $\phi^\alpha$, where the parameter $\alpha$ indexes the ensemble. It is useful to think of $\alpha$ taking a few integer values, \change{with the ensemble consisting of a few different flows}, or real values, so that the ensemble consists of a one-parameter family of flows. The theory is not restricted to these cases however and applies to the most general situation when the ensemble is determined by specifying a measure on $\SDiff$.

\begin{figure}
\begin{center}
\begin{tikzpicture}[scale=0.9]

\fill [ball color=white]
(0,0) .. controls +(0,2)  and +(-1,0)  .. (4,4) .. controls +(1,0) and +(0,2) .. 
(7,1.4) .. controls +(0,-2) and +(0,-2) .. (0,0);

\coordinate (A) at (1,0.5);
\coordinate (B) at (4,2);
\coordinate (C) at (3,3);
\coordinate (D) at (6,1.2);
\coordinate (E) at (3,-0.5);

\draw [fill=gray!80]  (A) circle [radius=0.06];
\node [below] at (A) {$x$};
\draw [fill=red]  (B) circle [radius=0.06];
\draw[>=stealth,shorten >=2pt,->,thick] (A) to[out=50,in=120] node[below] {$\bar \phi$} (B) ;
\draw [fill=gray!80]  (C) circle [radius=0.06];
\draw[>=stealth,shorten >=2pt,->] (A) to[out=70,in=120] node[below] {$\phi^1$} (C) ;
\draw [fill=gray!80]  (D) circle [radius=0.06];
\draw[>=stealth,shorten >=2pt,->] (A) to[out=-30,in=220] node[above right] {$\phi^2$} (D) ;
\draw [fill=gray!80]  (E) circle [radius=0.06];
\draw[>=stealth,shorten >=2pt,->] (A) to[out=-30,in=200] node[below left] {$\phi^3$} (E) ;

\draw[color=blue,>=stealth,shorten >=2pt,->] (B) to[out=80,in=-20] node[above] {$\xi^1$} (C) ;
\draw[color=blue,>=stealth,shorten >=2pt,->] (B) to[out=0,in=100] node[above] {$\xi^2$} (D) ;
\draw[color=blue,>=stealth,shorten >=2pt,->] (B) to[out=-100,in=50] node[left] {$\xi^3$} (E) ;

\node at (5,3) {$M$};


\end{tikzpicture}

\vspace{-3ex}
\caption{Decomposition $\phi^\alpha = \xi^\alpha \circ \bar \phi$ of the flow maps shown acting on a single point $x \in M$.}
\label{fig:dec1}

\end{center}
\end{figure}
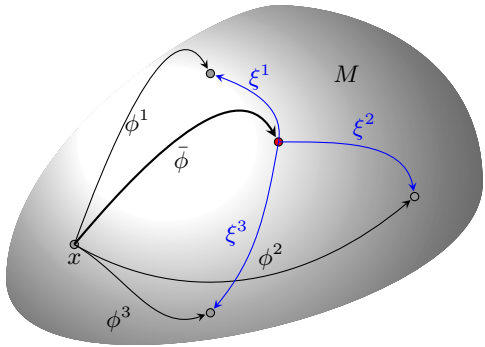 

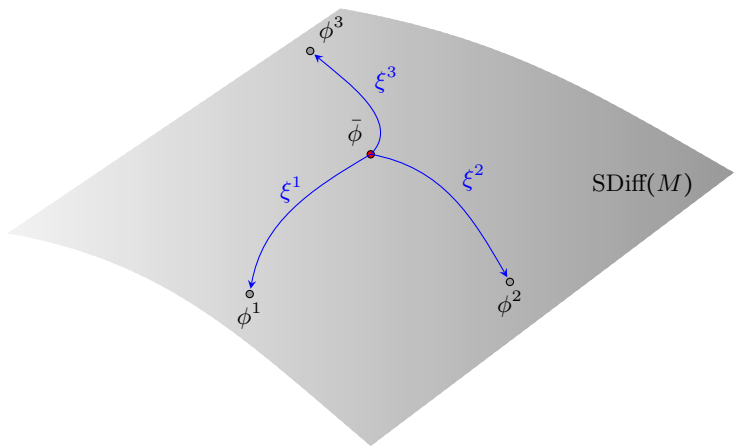
\begin{figure}

\begin{center}

\begin{tikzpicture}[scale=.8]

\shade[left color=gray!10,right color=gray!80] 
  (0,0) to[out=-10,in=140] (6,-3.5) -- (12,1) to[out=150,in=-10] (5.5,3.7) -- cycle;

\coordinate (A) at (6,1.3);
\coordinate (B) at (4,-1);
\coordinate (C) at (8.3,-.8);
\coordinate (D) at (5,3);

\draw [fill=red] (A) circle [radius=0.06];
\node [above left] at (A) {$\bar \phi$};

\draw [blue,>=stealth,shorten >=2pt,->] (A) to[out=210,in=80] node[above] {$\xi^1$} (B);
\draw [fill=gray!80]  (B) circle [radius=0.06];
\node [below] at (B) {$\phi^1$};

\draw [blue,>=stealth,shorten >=2pt,->] (A) to[out=-10,in=120] node[above right] {$\xi^2$} (C);
\draw [fill=gray!80]  (C) circle [radius=0.06];
\node [below] at (C) {$\phi^2$};

\draw [blue,>=stealth,shorten >=2pt,->] (A) to[out=45,in=-40] node[above right] {$\xi^3$} (D);
\draw [fill=gray!80]  (D) circle [radius=0.06];
\node [above right] at (D) {$\phi^3$};

\node at (10.5,.8) {$\textrm{SDiff}(M)$};


\end{tikzpicture}

\caption{Decomposition $\phi^\alpha = \xi^\alpha \circ \bar \phi$ of the flow maps, with each map thought of as a point in the infinite-dimensional group of diffeomorphisms $\SDiff$.}
\label{fig:dec2}
\end{center}
\end{figure}

It is natural to write each fluctuating flow map $\phi^\alpha$ in the ensemble as the composition
\beq \lab{meanpert}
\phi^\alpha = \xi^\alpha \circ \bar \phi
\eeq
of a perturbation map $\xi^\alpha$ with a mean flow map $\bar \phi$.  The maps $\phi^\alpha$ then stay close to $\bar\phi$ when
the maps $\xi^\alpha$ stay  close to the identity. Expression \eqn{meanpert} was used, for example, by \citet{Ho02a,Ho02b} and \citet{MaSh03}, and naturally extends to arbitrary manifolds the sum $\bx$ + $\bs{\xi}(\bx,t)$ of mean position $\bx$ and perturbation $\bs{\xi}(\bx,t)$ of standard GLM.  It can be visualised either by thinking of the decomposition of $\phi^\alpha x$ for a particular point $x \in M$ (see figure \ref{fig:dec1}), or by thinking of $\phi^\alpha$ as a point on the infinite-dimensional manifold $\SDiff$ (see figure \ref{fig:dec2}). Note that we do not insist that $\phi^\alpha = \id$ at $t=0$. The ensemble of maps can accommodate an ensemble of different initial conditions.

Of course the foregoing cannot uniquely define the mean map.
Nor is the problem of defining it straightforward, since we cannot use
linear averaging.  The definition has to be geometrically
intrinsic, thus coordinate independent, and preferably ensure that
$\bar\phi \in \SDiff$.  We discuss possible definitions in
\S\,\ref{sec:meandef}, and in \S\,\ref{sec:smallamp} examine their form under the simplifying
assumption of small-amplitude perturbations.  In the remainder of
the current section, we derive averaged dynamical equations that
hold for arbitrary $\bar\phi$.  This highlights properties that
depend only on the geometric structure of the Euler and related
equations and on the mean--perturbation decomposition (3.1),
independently of any particular definition of $\bar\phi$.

\subsection{Geometric GLM} \label{sec:geometricGLM}

We start by defining the mean velocity field $\baru$ as the time derivative of the mean flow map: 
\beq \lab{baru}
\baru = \dot{\bar \phi} \circ \bar \phi^{-1}.
\eeq
We emphasise that this is not the Eulerian-mean velocity: $\baru \not= \av{u^\alpha}$.  
While the notation $\bar u$ is often employed for the  Eulerian-mean velocity, we will reserve the explicit $\av{u^\alpha}$ for the few instances when this velocity will be needed. While our choice of notation is unconventional, it is completely natural in view of  the right-hand side of \eqn{baru} and we trust that it will not lead to confusion.  Our $\bar u$ is a Lagrangian mean velocity  in the sense that the flow map it generates takes fluid particles $a$ to positions $\bar \phi a$ that are as close as possible, in some sense, to the ensemble of
positions $\phi^\alpha a$.  We nonetheless refrain from using the notation $\barL{u}$ \change{or terming it `Lagrangian-mean velocity'} for reasons that will become apparent shortly. 

Taking the time derivative of \eqn{meanpert} and evaluating it at $x=\bar \phi a$ leads to
\beq \lab{key}
\dot \xi^\alpha x + ( \xi^\alpha_* \baru) (\xi^\alpha x) = u^\alpha (\xi^\alpha x), 
\eeq
with  $u^\alpha = \dot{\phi}{}^\alpha \circ (\phi^\alpha){}^{-1}$ as in \eqn{velocity1}. Defining the ensemble of perturbation velocity fields
\beq\lab{walphadef} 
w^\alpha = \dot \xi^\alpha \circ (\xi^{\alpha})^{-1},
\eeq
\eqn{key} can be written as an equality between vector fields, 
\beq \lab{uuw}
w^\alpha + \xi^\alpha_* \baru = u^\alpha.
\eeq
Pulling back by $\xi^{\alpha}$ and averaging leads to the equality
\beq \lab{baruuw}
\change{\baru = \xi^{\alpha*}(u^\alpha-w^\alpha) = \av{\xi^{\alpha*}(u^\alpha-w^\alpha)},} 
\eeq
\change{with the second equality following from $\av{\bar u} =\bar u$. This} can be thought of as a geometrically consistent version of the GLM equation $\bar{\bs{u}}^\mathrm{L}(\bx) = \av{\bu(\bx + \bxi(\bx))}$. On a general manifold $M$ the latter equality would involve averaging vectors lying in different tangent planes and so is simply not defined. Similarly, \eqn{uuw} indicates that $w^\alpha$ and $\xi^{\alpha*}w^\alpha$ are geometrically consistent relatives of GLM's $\bs{u}^\ell(\bx)=\bs{u}(\bx+\bxi(\bx))-\barL{\bu}(\bx)$.

The main aim of Lagrangian-mean theories is the derivation of equations governing the mean flow that take as simple a form as possible. For a perfect incompressible fluid, this is best achieved starting with the Euler equation \eqn{euler} in terms of the one-form $\nu^\alpha$ for each member $\alpha$ of the ensemble, pulling back with $\xi^\alpha$ then averaging. For \change{stratified Boussinesq} fluids, compressible fluids and MHD flows, a similar procedure can be applied to \eqn{boussinesq}, \eqn{compeulerouter} and \eqn{MHD1}, \eqn{MHD2}, respectively. A simple property, derived in \change{\citet{SoRo10,SoRo14} and in appendix \ref{sec:pullbackcons}}, is key to this procedure: for any ensemble- and time-dependent tensor $\tau^\alpha$,
\beq \lab{pullbackcons}
\xi^{\alpha*} \left(\partial_t  + \lie_{u^\alpha}  \right)  \tau^\alpha =  \left( \partial_t  + \lie_{\baru}  \right) (\xi^{\alpha*} \tau^\alpha).
\eeq
Note how the Lie derivative on the right-hand side is  taken with respect to the mean \change{velocity} $\baru$ which, however the mean flow map is chosen, is fluctuation-free, i.e.\ the same for each realisation  $\alpha$. Applying this to the Euler equation \eqn{euler}, using \eqn{usefulprops} and averaging yields
\beq \lab{avmom1}
(\partial_t + \lie_{\baru}) \av{\xi^{\alpha *} \nu^\alpha} = -  \, \d  \av{\xi^{\alpha *}  \pi^\alpha}.
\eeq
This suggests defining the Lagrangian mean of any tensorial object $\tau^\alpha$ as
\beq \lab{barL}
\barL{\tau}  = \av{\xi^{\alpha *} \tau^\alpha}.
\eeq
We emphasise the following: with this definition, the mean velocity $\bar u$ is not the Lagrangian mean of the velocity in the sense of the Lagrangian-mean operator \eqn{barL},
\beq \lab{barunot}
\baru \not= \barL{u}.
\eeq
For the equality to hold, that is to say for $\bar u = \av{\xi^{\alpha *} u^\alpha}$ to hold, we would have to impose that $\barL w = \av{\xi^{\alpha*} w} = 0$ (see \eqn{baruuw}).  However, as detailed in \S\,\ref{sec:meandef}, while this apparently provides a possible definition for \change{the mean flow map, it is unacceptable because it leads to a mean flow} that drifts away from the fluctuating flows, even if
the latter are wavelike and form a cluster in $\SDiff$. Note that standard GLM manages to maintain the equality in \eqn{barunot} but only by sidestepping intrinsic geometry (by transforming vector fields as $\bs{u}(\bs{x} + \bs{\xi}(\bs{x}))$, i.e.\ using parallel transport, rather than following proper pull-back transformation rules involving $\bs{u}(\bs{x} + \bs{\xi}(\bs{x})) \cdot (\nabla (\bs{x} + \bs{\xi}(\bs{x}))^{-1}$, with the superscript $-1$ denoting matrix inverse, as appropriate even in Euclidean spaces; see \citet{SoRo10}).

With the definition \eqn{barL}, $\barL{\nu} = \av{\xi^{\alpha *} \nu^\alpha}$ and \eqn{avmom1} takes the simple form
\beq \lab{aveuler}
\partial_t \barL{\nu} + \lie_{\baru} \barL{\nu} = - \,\d \barL{\pi}.
\eeq
This is a central result for any GLM-like theory: irrespective of the specific definition of the mean flow map $\bar \phi$, the Lagrangian-mean momentum, defined as a one-form, satisfies the averaged Euler equations \eqn{aveuler}. An Euclidean-space version was obtained by \citet{SoRo10}; the manipulations of \citet{AnMc78a}, following \citet{So72} and leading to their Theorem I, can similarly be interpreted as the \change{pull-back} of the momentum equation leading to \eqn{aveuler}.
Note  that \eqn{aveuler} can be derived from an averaged variational principle, in a manner that generalises to arbitrary manifolds the derivation of \citet{Sa13}. This is described in appendix \ref{sec:GLMlagr}.

A consequence of \eqn{aveuler} is that the Lagrangian-mean Kelvin circulation is conserved,
\beq
\oint_{\bar \phi C_0} \barL{\nu} = \textrm{const.},
\eeq
when the contour of integration is transported by the mean velocity $\baru$. Equivalently, the Lagrangian-mean vorticity
\beq
\d \barL{\nu} = \barL{\d \nu} 
\eeq
is conserved as a two-form: 
\beq
\partial_t \d \barL{\nu} + \lie_{\bar u} \d \barL{\nu} =0 . 
\eeq
Naturally, these also follow by applying $\xi^{\alpha *}$ to \eqn{Kelv} and \eqn{vorteqn}, and averaging. \change{Key to these calculations are the properties \eqn{usefulprops} that the operator $\d$ commutes with pull-backs and Lie derivatives.} 

The physics of the wave--mean flow interactions is encoded in the relation between the mean velocity $\bar u$ and Lagrangian-mean momentum $\barL \nu$, that is, between the advecting velocity and advected momentum in \eqn{aveuler}. Unlike in the original Euler equations, these are not related through the inertia operator $\flat$  associated with the metric $g$. Rather, they are related in a more complicated way, which depends on the precise definition of the mean flow map and on the transport of the metric by the $\xi^\alpha$, arising when the pull-back is applied to $\nu^\alpha$ as defined in \eqn{inertia}.

 It is natural to consider the difference
 \beq \lab{pmo}
 - \pmo = \barL{\nu} - (\bar u)_\flat
 \eeq 
between the Lagrangian-mean momentum and the one-form dual to the mean velocity vector since this characterises the manner in which waves or perturbations modify the mean Kelvin circulation, hence the mean vortical dynamics, compared with that of the Euler equations. The one-form $\pmo$ defined in \eqn{pmo} generalises the pseudomomentum $\textbf{\textsf{p}}$ of \citet{AnMc78a}, and we will use the same terminology. (Note that the sign of $\pmo$ is chosen according to a GLM standard convention ensuring that $\pmo$ reduces to the familiar product of wave action with wavevector in a small-amplitude, WKB regime; see \citealt{AnMc78a}). Parameterising the effect of perturbations on the mean flow can then be interpreted as providing an expression for $\pmo$, for example by means of a closure in terms of $\baru$.  

Expressions for $\pmo$ making it clear that it is quadratic to leading order in the perturbation amplitude $\eps$, say, (and is thus an $O(\eps^2)$ `wave property') are readily found from \eqn{pmo}. Defining the perturbation velocity $\tw^\alpha$ alternative to $w^\alpha$ by
\beq\lab{ualphaelldef}
\tilde w^{\alpha} = u^\alpha - \baru = w^\alpha + (\xi^{\alpha}_* - \id) \baru,
\eeq
we rewrite \eqn{pmo} as
\begin{subequations} \lab{pmos}
\begin{align}
- \pmo &  = \av{(\xi^{\alpha *} - \id) \baru_\flat} + \av{\xi^{\alpha*} \tw^\alpha_\flat} \\
& = \av{\xi^{\alpha *} ( \xi^\alpha_* \baru )_\flat - \baru_\flat }+ \av{ \xi^{\alpha*} w^\alpha_\flat}  . 
\lab{pmos1}
\end{align}
The right-hand sides are quadratic since, to leading order, $\xi^\alpha$ is the identity at $O(1)$ and the velocities $\tw^\alpha$ and $w^\alpha$ have vanishing averages at $O(\eps)$. 
The second term in each of these equivalent expressions is reminiscent of the standard GLM definition of pseudomomentum. 
Equation \eqn{pmos1} can also be written as 
\beq
- \pmo = \barL{g}  (\baru , \cdot) - \baru_\flat + \av{ \xi^{\alpha*} w^\alpha_\flat},  \lab{pmos2}  
\eeq
\end{subequations}
where $\barL{g} = \av{\xi^{\alpha *}  g }$, consistent with our general definition \eqn{barL} of Lagrangian averaging. The first two terms represent the effect on the mean flow of the wave-induced change in the average geometry; the third term represents the effect of the wave velocities, as in standard GLM.

\change{\subsection{Other fluid models} \label{sec:avother}}

\change{We now sketch the construction of mean equations for the more complex fluid models of \S\,\ref{sec:othermodels}. The Lagrangian-mean version of the three-dimensional Boussinesq equations \eqn{boussinesq} is obtained by applying $\xi^{\alpha *}$ to \eqn{boussimom}--\eqn{boussibuo} to obtain the momentum and buoyancy equations
\begin{subequations} \lab{barLboussi}
\begin{align}
\partial_t \barL{\nu}_\mathrm{a} + \lie_{\bar u}\barL \nu_\mathrm{a} &= - \d \barL \pi + \barL{\theta \, \d \Phi}, \lab{barLboussimom} \\
\partial_t \barL \theta + \lie_{\bar u} \barL \theta &= 0, \lab{barLboussibuo} 
\end{align}
\end{subequations}
and to \eqn{boussiPVcons} to obtain the conservation of the Lagrangian-mean potential vorticity
\beq
\left(\partial_t  + \lie_{\bar u} \right) \barL{\d \nu_\mathrm{a} \wedge \d \theta} = 0.
\eeq
The last term in \eqn{barLboussimom}, explicitly $\av{(\xi^{\alpha*} \theta^\alpha) \d (\xi^{\alpha*} \Phi^\alpha)}$,  is inconvenient but can be simplified by assuming that the buoyancy $\theta^\alpha$ in each flow realisation $\alpha$ results from the transport of the same initial buoyancy distribution, $\theta_0$ say. With this assumption,
\beq
\xi^{\alpha*} \theta^\alpha = \bar \phi_* \theta_0 = \barL \theta
\eeq
is independent of $\alpha$ and hence equal to its average.
The momentum and potential-vorticity equations then simplify as
\beq
\partial_t \barL{\nu}_\mathrm{a} + \lie_{\bar u}\barL \nu_\mathrm{a} = - \d \barL \pi + \barL{\theta} \d \barL \Phi 
\lab{boussiavmom}
\eeq
and
\beq
\left(\partial_t  + \lie_{\bar u} \right) \barL{\mathcal{Q}} = 0, \quad \textrm{where} \ \ \barL{\mathcal{Q}} = \d \barL \nu_\mathrm{a} \wedge \d \barL \theta.
\lab{boussiavpv}
\eeq
The Lagrangian-mean Rossby--Ertel potential vorticity $\barL{Q} = \av{\xi^{\alpha *}Q^\alpha}$ is also transported by the flow. When $\bar u$ is non-divergent (i.e., $\bar \phi$ preserves volume), it is related straightforwardly to its three-form counterpart by $\barL{Q} \omega = \barL{\mathcal{Q}}$. When $\bar u$ is divergent, this relationship is more complicated: introducing a mean density $\bar \rho$ such that $\bar \phi_* \omega = \bar \rho \omega$, it reads as $\barL{Q} \bar \rho \omega = \barL{\mathcal{Q}}$, noting that $\xi^{\alpha*} \omega = \bar \rho \omega$ is independent of $\alpha$.
The potential-vorticity equation is most useful in the limit of rapid rotation and strong stratification, when the mean flow can be assumed to be near geostrophic and in hydrostatic balance, and entirely slaved to the potential vorticity. We discuss this further in \S\ref{sec:Bouss}.}

 For a compressible fluid governed by \eqn{compeuler}, applying $\xi^{\alpha *}$ and averaging leads to 
\begin{subequations} \lab{compeulerav}
\begin{align}
\partial_t \barL{\nu} + \lie_{\bar u} \barL{\nu} &= - \, \barL {\rho^{-1} \d  p} + \tfrac{1}{2} \d \barL{g(u,u)} ,
 \lab{momcom} \\
\partial_t \barL{m} + \lie_{\baru} \barL{m} &= 0, \\
\partial_t \barL{\theta} + \lie_{\baru} \barL{\theta} &= 0, 
\end{align}
\end{subequations}
on using \eqn{pullbackcons}. 
The momentum equation \eqn{momcom} is complicated by a right-hand side that is not immediately expressed in terms of the Lagrangian mean of $p^\alpha$, $\rho^\alpha$ and $\nu^\alpha$. Furthermore, averaging the equation of state gives $\barL{p} = \av{P (\xi^{\alpha *} \rho^\alpha, \xi^{\alpha *} \theta^\alpha)}$, with a right-hand side that is, again, not expressed in terms of Lagrangian means. However, the conservation of the potential vorticity $\d \nu^\alpha \wedge \d \theta^\alpha$ \change{in \eqn{PV}} implies that the Lagrangian-mean potential vorticity 
\beq
(\partial_t  + \lie_{\baru}) \barL{\d \nu \wedge \d \theta} = 0
\eeq
is conserved. A conserved scalar in three dimensions is obtained \change{as the quotient of} the conserved three-forms $\barL{\d \nu \wedge \d \theta}$ and $\barL{m}$. 

\change{As in the Boussinesq case, simplifications} arise if we assume that the mass $m^\alpha$ and entropy $\theta^\alpha$ in each of the flow realisations $\alpha$ are obtained by transporting  single initial distributions, $m_0$ and $\theta_0$, say. In this case, 
\beq \lab{barLm}
\xi^{\alpha *} m^\alpha = \bar \phi_*  m_0 = \barL{m} \inter{and} \xi^{\alpha *} \theta^\alpha = \bar \phi_*  \theta_0 = \barL{\theta}
\eeq
are both independent of $\alpha$ and thus equal to their average. \change{(The first assumption is also made by \citet{BhFeMaMoWe05} in their Lagrangian averaging of compressible flows.)}
This simplifies the averaged equation of state into $\barL{p} = \av{P (\xi^{\alpha *} \rho^\alpha, \barL{\theta})}$ and, more importantly, the form of the conserved Lagrangian potential-vorticity volume form,
\beq
\barL{\d \nu \wedge \d \theta} = \d \barL{\nu} \wedge \d \barL{\theta},
\eeq
and of its scalar equivalent.

\change{
The averaging of the MHD equations parallels that of the Euler equations; see \cite{Ho02b} for GLM and related developments. Applying $\xi^{\alpha*}$ \change{to the induction equation \eqn{MHD1} and averaging gives}
\begin{equation}
(\partial_t + \lie_{\baru})  \barL B = 0. 
\end{equation}
The mean magnetic field $\bar{b}$ defined by $\iota_{\bar{b}} \omega = \barL{B}$  remains divergence free since $\d \barL{B} = 0$, even though the mean velocity $\baru$ may be divergent (this is equivalent to the pull-back of $b$ under $\xi^\alpha$ with tensor weight $-1$ in \cite{SoRo14}). To make some progress with the momentum equation we again assume that the mass distribution and magnetic flux are transported in each realisation from the same initial mass $m_0$ and flux $B_0$, so that 
\beq
\xi^{\alpha *} m^\alpha = \bar \phi_*  m_0 = \barL{m} 
\inter{and} 
\xi^{\alpha *} B^\alpha = \bar \phi_*  B_0 =  \barL{B}, 
\eeq
independent of $\alpha$. With $\bar{\rho} \omega = \barL{m}$ defining a mean density $\bar{\rho}$, we find that $\ip_{\bar b / \bar \rho} \, \barL m = \barL B$; since $\barL m$ and $\barL B$ are both transported by the flow, so is $\bar b / \bar \rho$ (as a vector field). {We note that
\beq \lab{Brho}
\xi^{\alpha*}(b/\rho)=\barL{b/\rho} = \bar b / \bar \rho
\eeq
follows from pulling back $ \ip_{b^\alpha/\rho^\alpha} m^\alpha = B^\alpha$ by $\xi^\alpha$  to obtain
\beq
{\ip_{\xi^{\alpha*}(b^\alpha/\rho^\alpha)} \,\xi^{\alpha*} m^\alpha} = {\ip_{\xi^{\alpha*}(b^\alpha/\rho^\alpha)}} \barL m =  {\xi^{\alpha*} B^\alpha} = \barL{B} = \ip_{\bar b / \bar \rho}\, \barL m
\eeq
and averaging.

Applying $\xi^{\alpha*}$ to \eqn{MHD2} written as 
\begin{align}
(\partial_t  + \lie_{u^\alpha} ) \nu^\alpha &=  \iota_{b^\alpha/\rho^\alpha}  \, \d \beta^\alpha  - (\rho^\alpha)^{-1} \d p^\alpha +  \tfrac{1}{2}  \d g(u^\alpha,u^\alpha),  
\lab{MHD2a}
\end{align}
averaging and using \eqn{Brho} yields
\begin{align}
(\partial_t  + \lie_{\baru} ) \barL{\nu} &=   \iota_{\bar{b}/\bar{\rho}}  \, \d \barL{\beta}  - \barL{(\rho^\alpha)^{-1} \d p^\alpha}  +  \tfrac{1}{2}  \d \barL{g(u^\alpha,u^\alpha) } . 
\lab{MHD2b}
\end{align}
It is then natural to define the one-form field $\hmo$ by, say,}
\beq \lab{hmo}
- \hmo = \barL{\beta} - (\bar{b})_\flat, 
\eeq 
which, like the pseudomomentum \eqn{pmo}, encodes the effect of the waves on the mean field in \eqn{MHD2b} and measures the lack of commutation between averaging and application of the inertia operator, in this instance to the magnetic field. This field is termed (minus) the `magnetisation induced by the Lagrangian mapping' by \cite{Ho02b}, but could also be called the `pseudomagnetising force' \citep[cf.][]{RoSo06a} or perhaps just the `pseudofield' for short. 

Finally suppose that the original MHD flow is incompressible, the density $\rho$ is a constant, and all the maps $\xi^\alpha$ are volume preserving (which excludes standard GLM). 
Then $\bar \rho$ also remains constant throughout and, taking this as unity, we can write 
\begin{subequations} \lab{MHDniceave}
\begin{align}
& (\partial_t + \lie_{\baru}) \bar{b}  = 0,\\
& (\partial_t  + \lie_{\baru}) \barL{\nu}  =  \iota_{\bar{b}}  \, \d \barL{\beta}  - \d \barL{\pi} =  \lie_{\bar{b}}  \, \barL{\beta}  - \d (\barL{\pi }+ \iota_{\bar{b}} \barL{\beta} ),  \\
& \divv \baru = 0 \quad \textrm{and} \quad \divv \bar{b} = 0 , 
\end{align}
\end{subequations}
without explicitly appealing to the two-form $B$.
}

\section{Definitions of the mean flow} \label{sec:meandef}

\subsection{Premiminaries} \label{sec:prelim}

In the previous section, we derived a generic form for the average of the Euler equations and their relatives and showed how key features, central to which is the simplicity and natural
averaging of Kelvin's circulation when expressed in terms of $\barL{\nu}$, hold regardless of the specific definition of the mean flow map. In this section we consider possible definitions of the mean flow map $\bar \phi$ from which the mean velocity $\bar u$ is deduced. 

As already discussed, there are two guiding principles in the choice of this definition. First, the mean flow map must be defined, if possible, in a way that ensures that the image of particles $\bar \phi a$ remains close to the images $\phi^\alpha a$ in each flow realisation, as one expects from the intuitive notion of mean. We expect this to be possible for
wavelike, as distinct from turbulent, flows.  `Remains close' can be made precise in the perturbative context: if the $\phi^\alpha$ are close together, in the sense that the maps ${(\phi^\alpha})^{-1} \circ \phi^\beta$ are close to the identity map for all $\alpha$ and $\beta$, then $(\phi^\alpha)^{-1} \circ \bar \phi$ should be close to the identity for long times (say times $O(\eps^{-1})$ or longer if the maps are $\eps$-close). 
Second, the definition of the mean should be geometrically sound in the sense that it does not depend on a particular coordinate choice and relies on the same geometric structure as the Euler equations themselves.

We remark that the first principle rules out simple definitions based on constraining the velocity perturbations $w^\alpha$ in \eqn{walphadef}, $\xi^{\alpha *} w^\alpha$ or $\tw^\alpha$  in \eqn{ualphaelldef} to have zero average. The constraints such definitions impose can ensure that the mean velocity is close to the  ensemble of velocities but they do not  prevent the drift of the mean trajectory away from the ensemble of trajectories over long times. 

To illustrate this point, we  consider a very simple set-up in which the velocities of the ensemble of flows consist of phase-shifted versions of a single, time-periodic, zero-mean velocity $U$: 
$u^\alpha(x,t)=\eps U(x, t - \alpha)$, where $\eps \ll 1$ and $\alpha \in \mathbb{R}$. Then, writing the perturbation map in coordinates as the expansion 
\beq \lab{xialphai}
\xi^{\alpha i}(x)   =  x^i + \eps  \xi_1^{\alpha i}(x) + \eps^2 \xi_2^{\alpha i}(x) + \cdots\, ,
\eeq
we  examine the implications for $\xi_2^{\alpha i}$ of constraining $w^\alpha$  or $\tw^\alpha$.
The perturbation velocity $w^\alpha$ associated with \eqn{xialphai} has the form
\beq
w^{\alpha i} = \eps \dot \xi_1^{\alpha i} + \eps^2 \left(\dot \xi_2^{\alpha i} - \xi_1^{\alpha j} \partial_j \dot \xi_1^{\alpha i}\right) + \cdots\, ,
\eeq
where $\dot \xi_1^{\alpha i} = U^{\alpha i}$. 
Imposing that $\av{w^\alpha}=0$, then gives 
\beq
\av{\dot \xi_2^{\alpha i}} =\av{ \xi_1^{\alpha j} \partial_j U^{\alpha i}},
\eeq
where the right-hand side, which can be recognised as a Stokes drift, does not vanish in general. This indicates that $\av{\xi_2^{\alpha i}}$ grows secularly and hence that the condition $\av{w^\alpha}=0$ does not lead to a satisfactory definition of the mean. To the order considered, it is equivalent to the use of the Eulerian mean which corresponds to $\av{\tw^\alpha}=0$, hence $\baru = \av{U}=0$ (see \eqn{ualphaelldef}), and whose deficiencies are well known. 

An alternative is to impose that $\av{\xi^{\alpha *} w^\alpha}= 0$, leading, as mentioned in \S\,\ref{sec:geometricGLM}, to the attractive feature that 
$\baru = \barL{u}$ \change{in the sense of \eqn{barL}}. However, this too is unsatisfactory, for the same reason. Indeed, a short calculation shows that
\beq
(\xi^{\alpha*} w^\alpha)^i =  \eps \dot \xi_1^{\alpha i} + \eps^2 \left(\dot \xi_2^{\alpha i}  - U^{\alpha j} \partial_j \xi_1^{\alpha i} \right) + \cdots\, ,
\eeq
so that
\beq
\av{\dot \xi_2^{\alpha i}} =\av{ U^{\alpha j} \partial_j \xi_1^{\alpha i}} , 
\eeq
which again indicates secular growth. 

These failures show that the definition of the mean flow map must
be based on constraining particle positions rather than particle
velocities. For this purpose we freeze the evolution of the flows and consider the ensemble of fluctuating flow maps $\phi^\alpha$ at a fixed time $t$. We want to define a mean flow map $\bar{\phi}$ that in some sense is as close as
possible to the maps $\phi^\alpha$. We consider several possibilities. Two are based on a distance between diffeomorphisms that is an alternative to the distance \eqn{D22} used to derive the Euler equations, namely
\change{\beq
\hat D^2(\phi,\psi) = \int_\M d^2(\phi,\psi) \omega 
= \inf_{\gamma_s: [0,1] \to \Diff} \int_0^1 \int_\M  g(q,q) \gamma_{s *} \omega \, \d s, \quad \gamma_0 = \phi, \ \ \gamma_1 = \psi, \lab{hD2}
\eeq
where $\gamma_s: [0,1] \to \Diff$ denotes an $s$-dependent diffeomorphism and 
$q =  \gamma_s' \circ \gamma^{-1}_s$
is the corresponding velocity field.}
Note that this distance  differs from \eqn{D22} in that it relaxes the assumption of volume preservation imposed on the maps $\gamma_s$ (even when the endpoint maps $\phi$ and $\psi$ are volume preserving).

We now discuss four possible definitions of the mean flow map $\bar\phi$ which we term `extended GLM', `optimal transport', `geodesic' and, following \cite{SoRo10}, `glm'. We keep the discussion at a formal level, ignoring for the most part the technical issues associated with the existence, smoothness and invertibility of $\bar \phi$ and only point out the most striking difficulties of some definitions in \S\,\ref{sec:remarks}. 

\subsection{Extended GLM} \label{sec:extGLM}

We define the mean flow as
\beq \lab{GLM}
\bar \phi = \argmin_{\phi \in \Diff} \av{\hat D^2(\phi,\phi^\alpha)},
\eeq
where $\argmin$ denotes the minimising map. This definition is coordinate-independent 
and can be recognised as the standard definition of the centre of mass on a Riemannian manifold, also termed  the Fr\'echet mean or Karcher mean (e.g., \citealt{Be03,Pe06}; see appendix \ref{app:frechet}). 
It will turn out -- see below \eqn{GLMconst} -- that this choice delivers a mean velocity field $\bar u$  that coincides with GLM's $\barL{\bu}$ when $M$ is Euclidean.

\change{The minimiser $\bar \phi$ is obtained by expanding the distance in \eqn{GLM} using \eqn{hD2} and 
seeking minimising paths $\gamma_s^\alpha$ joining $\bar \phi$ to $\phi^\alpha$ when $s$ runs from $0$ to $1$. This is best done using the associated $s$-dependent velocity fields 
\beq
q^\alpha = (\gamma^\alpha_s)' \circ (\gamma_s^\alpha)^{-1}, \lab{qalpha}
\eeq
where we recall that the prime denotes differentiation with respect to $s$. We stress that $s$ is a fictitious time variable: when measuring the distance between the two $t$-dependent diffeomorphisms  $\bar \phi$ and $\phi^{\alpha}$; the real time variable $t$ is frozen and treated as a fixed parameter for the paths $\gamma_s^\alpha$, $0\le s \le 1$. The vector field $q^\alpha$ associated with these paths is similarly unrelated to the velocities $\bar u$ and $u^{\alpha}$; it depends on both $s$ and $t$, though we do not make this dependence explicit. The perturbation maps $\xi^\alpha$ are simply given by $\gamma^\alpha_s$ at the endpoint $s=1$: since $\phi^\alpha = \gamma^\alpha_1 \circ \bar \phi$,
\beq \lab{qalpha1}
\xi^\alpha = \phi^\alpha \circ \bar \phi^{-1} = \gamma^\alpha_1.
\eeq}

A computation 
sketched in appendix \ref{sec:variother} shows that the one-forms $\qflat^\alpha =  g(q^\alpha,\cdot)$ satisfy the equation of a pressureless fluid in fictitious time, that is, 
\beq \lab{pressureless}
\partial_s \qflat^\alpha + \L_{q^\alpha} \qflat^\alpha =  \change{\tfrac{1}{2} \d g(q^\alpha,q^\alpha)},
\eeq
or equivalently,
\beq \lab{pressurelessq}
\partial_s q^\alpha + \nabla_{q^\alpha} q^\alpha = 0.
\eeq
\change{At the endpoints $s=0$ and $s=1$, the paths $\gamma^\alpha_s$ are constrained to satisfy $\gamma_0^\alpha=\bar \phi$ and $\gamma_1^\alpha=\phi^\alpha$, respectively. Minimisation over the $\alpha$-independent but otherwise arbitrary $\bar \phi$ imposes}
\beq \lab{pmean}
\av {\qflat^\alpha} = 0  \ \ \textrm{for} \ \ s = 0,
\eeq
\change{or, equivalently,
\beq \lab{GLMconst}
\av {q^\alpha} = 0 \ \ \textrm{for} \ \ s = 0,
\eeq 
since the inertia operator mapping $\qflat^\alpha|_{s=0}$ to $q^\alpha|_{s=0}$ is linear and $\alpha$-independent.} 

\change{The constraint \eqn{GLMconst}} determines the mean flow map implicitly. It can be made more transparent by noting that, for each $x \in M$ and each ensemble member $\alpha$, $\gamma_s^\alpha x$ is the geodesic on $M$ starting from $\bar \phi x$ at $s=0$ with speed $q^\alpha|_{s=0}$ to end at $\phi^\alpha x$ at $s=1$. \change{In the Euclidean case, the geodesics are straight lines, corresponding to $s$-independent vector fields $q^\alpha$ which can be identified with the displacement vectors $\bxi$ of standard GLM. Thus our construction directly generalises GLM's $\bs{x}^{\bxi}=\bs{x}+\bs{\xi}(\bs{x})$, with the constraint \eqn{GLMconst} generalising $\av{\bs{\xi}}=0$. It follows that, in this Euclidean case, the GLM velocity $\barL{\bu}$
corresponds to our mean velocity $\bar u$.  Note again
however that GLM's $\barL{\bu}$, our $\bar u$, is not a Lagrangian mean
in the sense of \eqn{barL}, for the reasons mentioned below \eqn{barunot}.
On curved manifolds, $\bar u$ is not obviously related to what may seem like a natural extension of \eqn{uLGLM}, namely the velocity obtained by parallel-transporting (using the Levi-Civita connection associated with the metric $g$, see \citet[][Ch.\ 9]{Fr97}) the velocities $u^\alpha$ along the geodesics $\gamma^\alpha_s x$ back to the mean position $x$, then averaging.

Our generalisation} of GLM to arbitrary manifolds $M$ respects their geometry in the sense that $\bar \phi$ maps $M$ to $M$ and it thus remedies the problem of mean trajectories leaving $\M$ encountered in standard GLM. Because there is no constraint that $\bar \phi$ be volume preserving, however, the mean velocity $\bar u$ is divergent in general \citep{Mc88}, an unavoidable price for the convenience of \eqn{GLMconst}. 
This can be addressed by including a volume-preservation constraint as we now describe.

\subsection{Optimal transport} \label{sec:OT}

An obvious generalisation of the above imposes volume preservation in the definition of $\bar \phi$ as a minimiser:
\beq
\bar \phi = \argmin_{\phi \in \SDiff} \av{\hat D^2( \phi,\phi^\alpha)}. \lab{opttran}
\eeq
This definition relates $\phi$ to the problem of optimal transportation \citep[e.g.,][]{Br03,Vi03}. Indeed, using \eqn{hD2}, the definition can be rewritten more explicitly as
\beq
\bar \phi = \argmin_{ \phi \in \SDiff}\,  \bigl\langle{ \,  \int_\M d^2( \phi ,\phi^\alpha ) \omega \, \bigr\rangle} . 
\eeq
Ignoring smoothness issues (which lead to a distinction between diffeomorphisms and more general rearrangements), 
we see that $\bar{\phi}$ is chosen among volume-preserving maps to ensure the minimum total cost for the transport from the mean fluid configuration to the perturbed configurations reached by the maps $\xi^\alpha = \phi^\alpha \circ \bar \phi^{-1}$.

Carrying out the minimisation in \eqn{opttran} leads, as before, to equation \eqn{pressureless} for a pressureless fluid. The constraint on $\bar \phi$ leads to the different end condition 
\beq
\av {\qflat^\alpha} = \d \psi \ \ \textrm{for} \ \ s = 0,
\lab{pmeanOT}
\eeq
where the function $\psi : M \to \mathbb{R}$ is a Lagrange multiplier enforcing the volume preservation of $\bar \phi$ (see appendix \ref{sec:variother}). 
The corresponding condition on the mean of the vector fields $q^\alpha$ now reads
\beq \lab{ceta1}
\av { q^\alpha} = \nabla \psi  \ \ \textrm{for} \ \ s = 0.
\eeq
\change{Eq.\ \eqn{ceta1} together with the requirement of geodesic paths  is obtained} by \citet{McC01} in his treatment of optimal transport on a manifold, here with additional averaging and without the technical condition of $\psi$ being $d^2/2$-convex (which is necessary to ensure that $\bar \phi$ is a minimiser and not just a critical point).

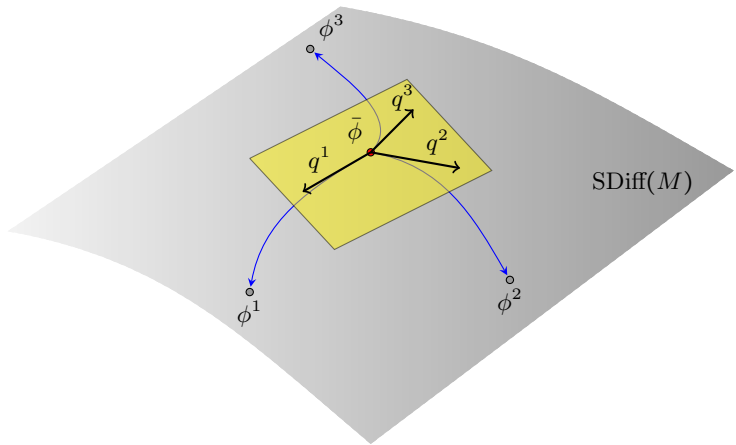
\begin{figure}
\begin{center}
\begin{tikzpicture}[scale=0.8]

\shade[left color=gray!10,right color=gray!80] 
  (0,0) to[out=-10,in=140] (6,-3.5) -- (12,1) to[out=150,in=-10] (5.5,3.7) -- cycle;

\coordinate (A) at (6,1.3);
\coordinate (B) at (4,-1);
\coordinate (C) at (8.3,-.8);
\coordinate (D) at (5,3);

\draw [>=stealth,shorten >=2pt,->,blue] (A) to[out=210,in=80] (B);
\draw [blue,>=stealth,shorten >=2pt,->] (A) to[out=-10,in=120] (C);
\draw [blue,>=stealth,shorten >=2pt,->] (A) to[out=45,in=-40] (D);

\draw [fill=yellow,opacity=0.5] (4,1.2)--(6.6,2.5)--(8,1)--(5.4,-0.3)--(4,1.2);

\draw [fill=red] (A) circle [radius=0.06];
\node [above left] at (A) {$\bar \phi$};


\draw [thick,>=latex,-to] (A) -- ++ (210:1.3) node[near end,above] {$q^1$};
\draw [fill=gray!80]  (B) circle [radius=0.06];
\node [below] at (B) {$\phi^1$};

\draw [thick,>=latex,-to] (A) -- ++ (-10:1.5) node[near end,above] {$q^2$};
\draw [fill=gray!80]  (C) circle [radius=0.06];
\node [below] at (C) {$\phi^2$};

\draw [thick,>=latex,-to] (A) -- ++ (45:1) node[near end,above] {$q^3$};
\draw [fill=gray!80]  (D) circle [radius=0.06];
\node [above right] at (D) {$\phi^3$};

\node at (10.5,.8) {$\textrm{SDiff}(M)$};


\end{tikzpicture}
\caption{Definition of $\bar \phi$ as Riemaniann centre of mass in the group of volume-preserving diffeomorphisms.}
\label{fig:geo}
\end{center}
\end{figure}

\subsection{Geodesic} \label{sec:geodesic}

\change{Since the distance \eqn{D22} on $\SDiff$ plays a fundamental role in incompressible fluid dynamics through the geodesic interpretation of the Euler equations,} the most natural average is arguably that defined as a Riemannian centre of mass using this distance:
\beq \lab{Sdiff}
\bar \phi = \argmin_{ \phi \in \SDiff} \av{D^2( \phi,\phi^\alpha)}.
\eeq
The maps $\phi^\alpha$ are reached from $\bar \phi$ (or equivalently the maps $\xi^\alpha$ are reached from $\id$) by integrating the time-dependent divergence-free velocity fields \AG{$q^\alpha = q^\alpha_s$} that satisfy the Euler equations in fictitious time $s$, 
\beq \lab{euler20}
\partial_s \qflat^\alpha + \L_{q^\alpha} \qflat^\alpha =  - \d \pi^\alpha,  \ \ \divv q^\alpha = 0,
\eeq
with the constraint that
\beq \lab{pmean2}
\av {\qflat^\alpha} = 0 \ \ \textrm{for} \ \ s = 0, 
\eeq
hence 
\beq \lab{ceta2}
\av { q^\alpha} = 0  \ \ \textrm{for} \ \ s = 0.
\eeq
Thus, compared with the extended GLM definition of \S\,\ref{sec:extGLM}, the only difference is the replacement of paths $\gamma_s^\alpha$ in $\Diff$ satisfying the pressureless fluid equations (and hence unconstrained by volume preservation) by paths in $\SDiff$ satisfying the Euler equations for incompressible fluids. Thus the construction of the mean flow map is taking place  entirely  within  $\SDiff$, as illustrated in figure \ref{fig:geo}. \change{Note that the presure $\pi^\alpha$ is also fictitious and is simply present to ensure that the flow $q^\alpha$ is divergence-free; we spare the reader the introduction of a new symbol here.} 

\change{While the mean-map definition \eqn{Sdiff} is inspired by the geodesic interpretation of the Euler equations, it can be used for other fluid models, even though  the distance on $\SDiff$ appearing in \eqn{Sdiff} is then no longer identical to the action functional governing the model's variational structure (for the rotating Boussinesq model, for instance, the action includes rotational and potential-energy terms; see \eqn{actionBouss}). In this case, the auxiliary one-form  $q^\alpha_\flat$ continues to satisfy the Euler equations \eqn{euler20} in contrast to the dynamical momentum}  $\nu^\alpha$ which satisfies more complicated equations. Alternatively, it may be possible to construct a mean flow map using only the relevant action functional, though sign-indefiniteness could make this problematic. We do not explore this possibility in the present paper.

\subsection{glm} \label{sec:glm}

The fourth possible definition of $\bar \phi$ we discuss is that proposed by \citet{SoRo10} in their `glm' reformulation of GLM. This relies on the group structure of $\SDiff$, but not on a metric. 
\change{In our framework, it simply amounts to choosing $q^\alpha$ to be divergence free and independent of the fictitious time $s$:
\beq \lab{glmcond}
\partial_s q^\alpha = 0 \ \ \textrm{with} \ \ \divv q^\alpha = 0.
\eeq
The perturbation maps can then be written as
\beq \lab{glm}
\xi^\alpha = \exp q^\alpha,
\eeq
where the exponential is the Lie-group exponential on $\SDiff$,    
defined as the flow map} at time $s=1$ of the $s$-independent \change{vector field} $q^\alpha$. 
The \change{glm mean flow map} is then defined by imposing
\beq \lab{glmmean}
\av{q^\alpha} = 0 .
\eeq
This definition is clearly coordinate independent and guarantees that the mean and perturbation maps are volume preserving. This is achieved by imposing the constraint of vanishing average to linear objects, the vector fields $q^\alpha$ which act as proxies for the perturbation flow maps $\xi^\alpha$. \change{(Note that we use the term `glm' for \citeauthor{SoRo10}' choice \eqn{glm}--\eqn{glmmean} without implying a restriction to small-amplitude perturbations that accompanied \citeauthor{Ho02b}'s (\citeyear{Ho02b}) original introduction of the term.)}

\subsection{Remarks} \label{sec:remarks}

We conclude this section with some remarks, first on the limitations of the definitions. 
It is clear that the four definitions can at best define a mean flow map locally, when the maps $\phi^\alpha$ are close  enough to one another. Any theory can at best work for finite amplitude displacements from the mean, but within finite limits which we will not attempt to specify here. For the first three, it is enough to note that even in  finite dimension, centres of mass often fail to exist globally, as the example of the mean of points uniformly distributed on the sphere demonstrates. Furthermore, it is possible, even in standard GLM in Euclidean geometry, that the mean flow map $\bar \phi$ fail to be invertible. Non-smooth solutions are usual in the theory of optimal transport and, while they may provide acceptable solutions in other contexts, it is unclear what their relevance would be for wave--mean-flow interaction. On the other hand, results on the local existence of smooth solutions  could be used to check that the definitions of the mean flow map make sense locally. 

For the geodesic definition of \S\,\ref{sec:geodesic}, in particular, the existence of a unique geodesic between sufficiently close volume-preserving diffeomorphisms \citep{EbMa70} suggests the local existence of a unique Riemannian centre of mass.  
\change{For the glm definition, given an ensemble of  fields $q^\alpha$ satisfying the conditions \eqn{glmcond}, a mean flow map $\bar \phi$ and corresponding perturbed maps $\phi^\alpha$ can be generated, and the governing equations can be averaged as, for example, in dynamo applications \citep{SoRo14}. If, however, we regard arbitrary perturbed maps $\phi^\alpha$ as a starting point, the glm definition is more problematic. This is because the set of flow maps that can be written as the exponential of an $s$-independent vector field in the manner of \eqn{glm} is limited (see \citealt{EbMa70}, \S9, for the difference between accessibility by geodesics and accessibility by the Lie-group exponential). To see this, it is enough to notice that the flows generated by time-independent vector fields cannot have isolated periodic points (since the whole orbit of such points consists of periodic points), and so cannot generate general maps.}

The constructions in \S\S\,\ref{sec:OT}--\ref{sec:glm} ensure that the mean flow map is volume preserving and are appropriate for the \change{Boussinesq} or incompressible MHD equations as well as for the incompressible Euler equations. For a compressible fluid, in contrast, there is no reason to insist that the mean flow map be volume preserving. It may be advantageous to impose other constraints that lead to simplifications of the mean dynamics. We suggest two possibilities. The first is to impose that the Lagrangian mean density $\barL \rho = \av{ \xi^{\alpha *} \rho^\alpha}$ and the mean mass form in \eqn{barLm} satisfy the natural relation $\barL m = \barL \rho \omega$.  Pulling back by $\bar \phi^*$, this constraint can be rewritten as $\av{\phi^{\alpha*} m^\alpha}=\av{\phi^{\alpha *} \rho} \, \bar \phi^* \omega$ and, since \eqn{barLm} implies that $\av{\phi^{\alpha *} m^\alpha} \, \av{\phi^{\alpha *} (\rho^{\alpha})^{-1}}=\av{\phi^{\alpha*} \omega}$,
\beq
\av { \phi^{\alpha *} \omega} = \av {\phi^{\alpha  *} \rho^\alpha}\,  \av {\phi^{\alpha *} (\rho^\alpha)^{-1}} \, {\bar \phi}^* \omega . 
\eeq
An alternative is to choose $\bar \phi$ to simplify the averaging of the equation of state. Writing this equation as $m^\alpha = R(p^\alpha,\theta^\alpha) \omega$, simplifications would ensue if $\bar \phi$ could be chosen to ensure that $\barL{m} = R (\barL p, \barL \theta) \omega$ also; the mean flow would then satisfy the same equation of state as the individual flows.

\section{Small-amplitude perturbations} \label{sec:smallamp}

\subsection{Preliminaries}

The definitions of the mean flow map in \S\S\,\ref{sec:extGLM}--\ref{sec:glm} are somewhat involved. 
\change{To exploit them in practice requires us to work perturbatively, order-by-order in a small amplitude parameter $\eps \ll 1$. This is conveniently done in terms of the $s$-dependent vector fields $q^\alpha$: the perturbative framework is introduced simply by expanding these in powers of $s$,
\beq
q^\alpha=q^\alpha_1 + s q^\alpha_2 +  \cdots,
\lab{qpert}
\eeq
where the $q^\alpha_i$ are independent of $s$, and by taking the maps $\xi^\alpha$ as the flows of $q^\alpha$ at $s=\eps$ instead of $s=1$. By repeated differentiation with respect to $s$ of the identity
\beq
\dt{}{s} ( \xi^*_s \tau) = \xi_s^* \lie_{q_s} \tau,
\eeq
where $\tau$ is an arbitrary tensor, we obtain the Taylor expansion
\beq
\xi^{\alpha *}\tau 
= \tau + \eps \lie_{q_1^\alpha} \tau + \tfrac{1}{2} {\eps^2} (\lie_{q_1^\alpha}^2 + \lie_{q_2^\alpha}) \tau + \cdots.
\lab{pullbackexp}
\eeq
In particular, applying this to the coordinate function $x^i$ gives
\beq
\xi^{\alpha i}(x) = x^i + \eps \xi_1^{\alpha i} +  \eps^2 \xi_2^{\alpha i} + \cdots = x^i + \eps q_1^{\alpha i} + \tfrac{1}{2} {\eps^2} \left(q_2^{\alpha i} + q_1^{\alpha j} \partial_j q_1^{\alpha i} \right) + \cdots.
\lab{xiq}
\eeq

Introducing \eqn{qpert} into the equation satisfied by $q^\alpha$ in each mean-flow definition leads to
\begin{subequations} \lab{q2}
\beq
\textrm{GLM, optimal transport:} \qquad q_2^\alpha = - \nabla_{q_1^\alpha} q_1^\alpha, \lab{q2OT}
\eeq
\beq
\textrm{geodesic:} \qquad \qquad q_2^\alpha = - \P \nabla_{q_1^\alpha} q_1^\alpha, \lab{q2geo}
\eeq
where $\P$ denotes projection onto divergence-free vector fields, and
\beq
\textrm{glm:} \qquad \qquad q_2^\alpha = 0.
\eeq
\end{subequations}
The constraint at the endpoint $s=0$ simply implies that $\av{q_1^\alpha}=0$ for the GLM, geodesic and glm definitions. For optimal transport, this only holds to leading order in $\eps$: the $O(\eps)$ function $\psi$ in \eqn{ceta1} is determined by the volume-preservation constraint $\bar \phi^* \omega = \omega$, hence $\av{\xi^{*\alpha} \omega}=\omega$; using \eqn{pullbackexp} this imposes that 
\beq
\divv \av{ q_1^\alpha + \tfrac{1}{2} {\eps} q_2^\alpha } = O(\eps^2) . 
\eeq
Since $\divv q_1=O(\eps)$ and, from \eqn{q2OT}, $q_2$ is divergent, the $O(\eps)$ part of $\av{q_1^\alpha}$ introduced by the gradient in \eqn{ceta1} must be chosen such that
\beq
\textrm{optimal transport:} \qquad
\av{q_1^\alpha + \tfrac{1}{2} {\eps} q_2^\alpha} =- \tfrac{1}{2} {\eps} \P \av{\nabla_{q_1^\alpha} q_1^\alpha} + O(\eps^2).
\lab{OTconstpert}
\eeq

\subsection{Mean flow and Lagrangian-mean momentum}

We can now express the  mean velocity $\bar u$, Lagrangian-mean momentum $\barL \nu$, and hence pseudomomentum $\pmo$ perturbatively in terms of the $q_i^\alpha$. We assume that the velocity and momentum have the dual expansions 
\beq
u^\alpha = u_0 + \eps u^\alpha_1 + \eps^2 u_2^\alpha + \cdots \inter{and} \nu^\alpha = \nu_0 +\eps \nu^\alpha_1 + \eps^2 \nu^\alpha_2 + \cdots,
\eeq
with $\av{u^\alpha_1}=\av{\nu^\alpha_1}=0$. 
The mean velocity is deduced from \eqn{baruuw}, which requires the expansion
\beq
w^\alpha = \eps \partial_t q^\alpha_1 + \tfrac{1}{2} {\eps^2} (\partial_t q^\alpha_2 - \lie_{q^\alpha_1} \partial_t q^\alpha_1) + \cdots.
\lab{wpert}
\eeq
This is obtained by considering the paths $\gamma_{s}^\alpha$ in \eqn{qalpha}--\eqn{qalpha1} as functions of both $s$ and $t$. Defining the corresponding velocities $w^\alpha_s =  \dot \gamma^{\alpha}_s \circ (\gamma_s^\alpha)^{-1}$ 
(such that $w^\alpha=w^\alpha_s|_{s=\eps}$), we can apply \eqn{liecommute} to find
\beq
\partial_s w_s^\alpha = \partial_t q_s^\alpha - \lie_{q^\alpha} w_s^\alpha , 
\eeq
from which \eqn{wpert} follows by Taylor expansion. 
 
Now, for the GLM, geodesic and glm mean flows, we can treat $q_1^\alpha$ and $q_2^\alpha$ as $\eps$-independent. This is not true for the optimal-transport mean flow since \eqn{OTconstpert} implies that $q_1^\alpha$ contains an $O(\eps)$ term. We handle this complication with minimal inconvenience by noting that, to the order $O(\eps^2)$ considered here, \eqn{pullbackexp} and \eqn{wpert} involve $q_2^\alpha$ only in the combination $q_1^\alpha + \tfrac{1}{2} \eps q_2^\alpha$. This makes it possible to redefine $q_1^\alpha$ as the value of $q^\alpha_1$ at $\eps=0$ and to incorporate its $O(\eps)$-part in a redefined $q_2^\alpha$. This reduces \eqn{OTconstpert} to $\av{q_2^\alpha}=-\P \av{\nabla_{q_1^\alpha}q_1^\alpha}$. Because this also holds for the geodesic mean flow, see \eqn{q2geo}, and because  $\av{q_2^\alpha}$ is the only part of $q_2^\alpha$ that matters to order $O(\eps^2)$, the form of $\bar u$ and $\barL \nu$ is the same for the optimal-transport and geodesic mean-flow definitions. We take advantage of this to bypass the complication associated with the optimal-transport definition in what follows. 

Introducing \eqn{wpert} into \eqn{uuw} gives to the first three orders $\bar u_0 = u_0$, 
\beq
\partial_t q^\alpha_1 + \lie_{u_0} q^\alpha_1 = u^\alpha_1
\lab{upert01}
\eeq
and, upon averaging,
\beq
\quad \bar u_2 = \av{u_2^\alpha} +  \tfrac{1}{2} \av{\lie_{q^\alpha_1} u_1^\alpha} - \tfrac{1}{2} \av{( \partial_t + \lie_{u_0}) q_2^\alpha},
\lab{barupert}
\eeq
where we have used that $\av{q^\alpha_1}=0$ as follows from \eqn{upert01}. 

The Lagrangian-mean momentum is computed from its definition: using  \eqn{pullbackexp} we find that
\beq
\xi^* \nu^\alpha = \nu_0 + \eps (\nu^\alpha_1 + \lie_{q^\alpha_1} \nu^\alpha_0) + \eps^2 ( \nu^\alpha_2 + \lie_{q^\alpha_1} \nu_1 + \tfrac{1}{2} \lie_{q^\alpha_1}^2 \nu_0 + \tfrac{1}{2} \lie_{q^\alpha_2} \nu_0) + \cdots .
\eeq
Averaging and using that $\av{q^\alpha_1}=0$, we obtain
\beq
\barL \nu = \nu_0 + \eps^2 \barL \nu_2 + \cdots,
\eeq
where
\beq
\barL \nu_2 =  \av{\nu_2^\alpha} + \av{\lie_{q^\alpha_1} \nu^\alpha_1} + \tfrac{1}{2} \av{\lie_{q^\alpha_1}^2 \nu_0} + \tfrac{1}{2} \av{\lie_{q^\alpha_2} \nu_0}.
\lab{barnupert}
\eeq

Eqs.\ \eqn{barupert} and \eqn{barnupert} give explicit, coordinate-free expressions for the first non-trivial term in the expansion of the mean velocity and Lagrangian-mean momentum. Using \eqn{q2}, we can specialise these expressions for each of the mean-flow definitions to obtain:
\begin{subequations}
\lab{u2nu2}
\begin{align}
\textrm{GLM:} \quad \bar u_2 &= \av{u_2^\alpha} + \tfrac{1}{2} \av{\lie_{q^\alpha_1} u_1^\alpha} + \tfrac{1}{2} ( \partial_t + \lie_{u_0}) \av{\nabla_{q^\alpha_1} q_1^\alpha}, \\
\barL \nu_2 &=  \av{\nu_2^\alpha} + \av{\lie_{q^\alpha_1} \nu^\alpha_1} + \tfrac{1}{2} \av{\lie_{q^\alpha_1}^2 \nu_0} - \tfrac{1}{2} \lie_{\av{\nabla_{q^\alpha_1} q_1^\alpha}} \nu_0, \\
\textrm{geodesic,  optimal transport:}  \quad \bar u_2 &=  \av{u_2^\alpha} + \tfrac{1}{2} \av{\lie_{q^\alpha_1} u_1^\alpha} + \tfrac{1}{2} ( \partial_t + \lie_{u_0}) \P \av{\nabla_{q^\alpha_1} q_1^\alpha}, \\
\barL \nu_2 &=  \av{\nu_2^\alpha} + \av{\lie_{q^\alpha_1} \nu^\alpha_1} + \tfrac{1}{2} \av{\lie_{q^\alpha_1}^2 \nu_0} - \tfrac{1}{2} \lie_{\P \av{\nabla_{q^\alpha_1} q_1^\alpha}} \nu_0, \\
\textrm{glm:} \quad \bar u_2 &=  \av{u_2^\alpha} + \tfrac{1}{2} \av{\lie_{q^\alpha_1} u_1^\alpha}, \\
\barL \nu_2 &=  \av{\nu_2^\alpha} + \av{\lie_{q^\alpha_1} \nu^\alpha_1} + \tfrac{1}{2} \av{\lie_{q^\alpha_1}^2 \nu_0}.
\end{align}
\end{subequations}
These are key results of this paper. Together with the Lagrangian-averaged momentum equation  \eqn{aveuler}, they govern the mean dynamics, providing an evolution equation for $\av{\nu_2^\alpha}$. The Eulerian-mean momentum  $\av{\nu_2^\alpha}$ appears as part of the advected Lagrangian-mean momentum $\barL \nu$, while the Eulerian-mean velocity $\av{u_2^\alpha}$ appears as part of the advecting mean velocity $\bar u$. They are explicitly related through any of the pairs of equations in \eqn{u2nu2}, depending on the choice of mean-flow definition, and by the identity $\av{\nu_2^\alpha} = \av{u_2^\alpha}_\flat$. The equation for $\av{\nu_2^\alpha}$ is not closed: a model for the dynamics of the ensemble of perturbations, linking $q_1^\alpha$ or $u_1^\alpha$ to the mean quantities (and consistent with \eqn{upert01}), is necessary to evaluate the quadratic perturbation terms in \eqn{u2nu2}. This can be done in a heuristic manner, as in the derivation of $\alpha$- and related models \citep[e.g.][]{Ho02b,MaSh03}, or based on the $O(\eps)$ dynamical equations. We do not consider this closure problem further in this paper.

It is interesting to express the condition satisfied by $\av{q_2^\alpha}$ in terms of the coordinate-dependent $\xi_2^{\alpha i}$ to connect our results with standard GLM. From \eqn{xiq}, \eqn{q2OT} and the form of the covariant derivative we find that
\beq
\textrm{GLM:} \qquad \av{\xi^{\alpha i}_2} = \tfrac{1}{2}  \av{q_1^{\alpha j} \partial_j q_1^{\alpha i}} -  \tfrac{1}{2}  \av{\nabla_{q_1^\alpha} q_1^{\alpha}}^i = - \tfrac{1}{2} \Gamma^{i}_{jk} \av{q_1^{\alpha j} q_1^{\alpha k}}.
 \lab{GLMmeanxi2}
\eeq
In the case where $M = \mathbb{R}^n$ with Cartesian coordinates, $\Gamma^{i}_{jk}=0$ and the standard GLM condition $\av{\xi_1^{\alpha i}}=\av{\xi_2^{\alpha i}}=0$ is recovered. On general manifolds, however, $\av{\xi^{\alpha i}_2}\not= 0$  as is required to ensure that the mean flow is tangent to the manifold. The mean flow is not volume preserving, however, since $\divv \av{q^\alpha} \not=0$. This is remedied for the other mean flows for which we obtain similarly
\beq  
\textrm{geodesic, optimal transport:} \qquad \av{\xi_2^{\alpha i}} = \tfrac{1}{2} \av { q_1^{\alpha j} \partial_j q^{\alpha i}_1} - \tfrac{1}{2} \P \av{\nabla_{q_1^\alpha} q_1^{\alpha}}^i \lab{xi2geo}
\eeq
and
\beq  
\textrm{glm:} \qquad \av{\xi_2^{\alpha i}} = \tfrac{1}{2} \av { q_1^{\alpha j} \partial_j q^{\alpha i}_1}.  \lab{xi2glm}
\eeq
The latter expression recovers a key result in \cite{SoRo10}. Observe the similarities between expressions \eqn{GLMmeanxi2}, \eqn{xi2geo} and \eqn{xi2glm} for $\av {\xi^{\alpha i}_2}$ obtained for the various definitions of the mean: the glm term \eqn{xi2glm} is common to all expressions; the other three expressions contain an additional term that depends on the metric and is rendered divergence free by projection in the case of the optimal transport and geodesic definitions. For $M = \mathbb{R}^n$, $\nabla_{\xi_1^\alpha} \xi_1^{\alpha i} = \xi_1^{\alpha j} \partial_j \xi_1^{\alpha i}$, and the difference between glm on the one hand, and optimal transport and geodesic on the other hand, is the projection $(\I - \P)$ on gradient vector fields (spanning the complement to divergence-free vector fields); thus, in this sense, $\av{\xi_2^{\alpha i}}$ is larger for glm than for the optimal transport and geodesic definitions of the mean.
}

\section{Application: wave--mean flow interaction in Boussinesq fluid}\label{sec:Bouss}

We illustrate the theory developed in this paper with a perturbative treatment of the Lagrangian-mean dynamics of a Boussinesq fluid derived in \S\ref{sec:avother}. We focus on the rapid-rotation, strong-stratification regime, where the mean flow is balanced, hence controlled by potential vorticity, while the perturbations represent fast inertia-gravity waves. Averaging is then naturally over the fast time scale of the waves. 

Our aim is to relate the transported Lagrangian potential vorticity $\barL{\mathcal{Q}}$ in \eqn{boussiavpv} to the advecting velocity $\bar u$ under quasi-geostrophic scaling. We make the `strong-wave' assumption $u_0=\nu_0=0$. This leads to $O(\eps^2)$ mean quantities and reduces \eqn{upert01}, \eqn{barupert} and \eqn{barnupert} to
\beq
\partial_t q_1^\alpha = u_1^\alpha, \quad \bar u_2^\alpha = \tfrac{1}{2} \av{\lie_{q_1^\alpha} u_1^\alpha} \inter{and}  \barL \nu_2 = \av{\lie_{q_1^\alpha} \nu_1^\alpha},
\eeq
using that $\av{\partial_t \,\cdot\, }=0$.
Correspondingly, the leading-order pseudomomentum is given by
\beq
- \pmo =  \av{\lie_{q_1^\alpha} \nu_1^\alpha} - \tfrac{1}{2} \av{\lie_{q^\alpha_1} u_1^\alpha}_\flat  ,
\eeq
and is identical for all mean-flow definitions. In components and in the Euclidean case on which we focus this is just
\beq
- \pmo_i = \av{\nu_j^\alpha \partial_i q_1^{\alpha j}}, \lab{pmostandard}
\eeq
again since $\av{\partial_t \,\cdot\, }=0$, consistent with standard results \citep{Bu14}. Note that we retain $\pmo = \bar u_\flat - \barL \nu$ as definition of the pseudomomentum, when we could equally well have included background-rotation effects by changing $\barL \nu$ to $\barL \nu_\mathrm{a}$.

We now consider the Lagrangian-mean momentum equations \eqn{boussiavmom} in the quasi-geostrophic regime. 
We use the familiar notation: $(x,y,z)$ for Cartesian coordinates, $(u,v,w)=(u^{\alpha 1},u^{\alpha 2},u^{\alpha 3})$ for the velocity components, and $(\xi,\eta,\zeta)=(\xi^{\alpha 1},\xi^{\alpha 2},\xi^{\alpha 3})$ for the displacement components, dropping the superscript $\alpha$ in the process. The dual use of $u$ and $\xi$ for a vector and its first component should not lead to confusion.
Taking $\Phi=z$ and $\theta_0(z)=N^2 z$, with constant $N$ for simplicity, \eqn{boussiavmom} reduces to the geostrophic and hydrostatic balances
\begin{subequations}
\begin{align}
- f \bar v &= - \barL \pi_x + \barL \theta \av{\zeta_2}_x, \lab{u} \\
 f \bar u &= - \barL \pi_y + \barL \theta \av{\zeta_2}_y,\lab{v} \\
 0 &= - \barL \pi_z + \barL \theta (1+\av{\zeta_2}_z), \lab{pressure}
\end{align}
\end{subequations}
where the average of the second-order vertical displacement $\zeta_2$ arises as a result of the pull-back of $\d \Phi = \d z$ \citep[cf.][]{XiVa15}. Note how the mean velocity components $\bar u$ and $\bar v$ appear on the left-hand side, and not the corresponding components of the momentum $\barL \nu$. The Lagrangian-mean buoyancy can be written as
\beq
\barL \theta = N^2 z + N^2 \av{\zeta_2} + \bar \psi_z/f,
\eeq
where $\bar \psi$ is an $O(\eps^2)$ streamfunction. Substituting into \eqn{pressure} and retaining only the $O(1)$ and $O(\eps^2)$ terms leads to
\beq
\barL \pi = \tfrac{1}{2} N^2 z^2 + N^2 z \av{\zeta_2} + \bar \psi/f. 
\eeq
Introducing this into \eqn{u}--\eqn{v} gives the geostrophic balance
\beq
\bar u = - \bar \psi_y \inter{and} \quad \bar v = \bar \psi_x. 
\eeq

We compute $\barL{\mathcal{Q}}$ from its definition \eqn{boussiavpv}, using that 
\beq
\barL \nu_\mathrm{a} = \av{\xi^{\alpha*} A}  +\bar u \, \d x + \bar v \, \d y - \pmo,
\eeq
where only the first term is $O(1)$. Using \eqn{pmostandard}, we obtain that 
\beq
-\d \pmo \wedge \d z = \av{\partial(u_1,\xi_1) + \partial(v_1,\eta_1)} \, \d x \wedge \d  y \wedge \d z
\eeq
where $\partial(\cdot,\cdot)$ is the Jacobian in $(x,y)$. 
Also,
\beq
\d \av{\xi^{\alpha *} A} = f \av{\xi^{\alpha *} (\d x \wedge \d  y)} = f  \left( 1 +  \av{\partial(\xi_1,\eta_1)} + \av{\xi_2}_x + \av{\eta_2}_y \right) \d x \wedge \d  y.
\eeq
Introducing into \eqn{boussiavpv} and using the quasi-geostrophic scaling to retain only the leading-order terms, we obtain
\begin{align}
\barL{\mathcal Q} = N^2 & \left( f  + \left(\partial_{xx} + \partial_{yy}  + f^2 N^{-2} \partial_{zz}\right) \bar \psi \right. \nonumber \\
& \left. + \av{\partial(u_1,\xi_1) + \partial(v_1,\eta_1)}  + f \av{\partial(\xi_1,\eta_1)} + f \av{\partial_i \xi_2^{\alpha i}} \right) \d x \wedge \d  y \wedge \d z.
\lab{aa}
\end{align}
For the optimal-transport, geodesic and glm mean-flow definitions, \eqn{xiq} and the property $\partial_i q_2^i=0$ resulting from  the volume-preservation of $\xi^\alpha$ imply that
\beq
\av{\partial_i \xi_2^{\alpha i}}  = \tfrac{1}{2}  \partial_i  \av{\xi_1^{\alpha j} \partial_j \xi_1^{\alpha i}} = \tfrac{1}{2} \partial_{ij} \av{\xi_1^{\alpha i} \xi_1^{\alpha j}}.
\eeq
The material conservation of $\barL{\mathcal{Q}}=\barL Q \omega$ can finally be written as
\beq
\partial_t \barL Q + \partial(\bar\psi,\barL Q)=0, 
\lab{QGPVcons}
\eeq
where
\beq
\barL Q = \left(\partial_{xx} + \partial_{yy} + \frac{f^2}{N^2} \partial_{zz} \right) \bar \psi+\av{\partial(u_1,\xi_1) + \partial(v_1,\eta_1)}+f\av{\partial(\xi_1,\eta_1)} + \tfrac{1}{2} {f} \partial_{ij} \av{\xi^{\alpha i}_1 \xi^{\alpha j}_1}.
\lab{QGPV}
\eeq
Eqs.\ \eqn{QGPVcons}--\eqn{QGPV} have been obtained by \citet{HoBuFe2011}, \citet{WaYo15} and \citet{Sa16}  using, respectively, standard GLM, multi-scale asymptotics, and a variational approach. They describe the advection of the Lagrangian-mean quasi-geostrophic potential vorticity which includes, in addition to the familiar (scaled) Laplacian of the streamfunction $\bar \psi$, a wave-induced contribution that is quadratic in the wave fields.

The derivation of \eqn{QGPVcons}--\eqn{QGPV} in the GLM framework is somewhat different.  In this case, the terms $\av{\xi^{\alpha i}_2}$ disappear from $\barL{\mathcal Q}$ in \eqn{aa}, but there is a compensating effect stemming from the divergence of $\bar u$. As noted in \S\ref{sec:avother}, $\barL Q$ is obtained from $\barL{\mathcal{Q}}$
by dividing by the density $\bar \rho$ such that $\bar \phi_* \omega = \bar \rho \omega $, and
given by $\bar \rho = 1 -\tfrac{1}{2} \partial_{ij} \av{\xi^{\alpha i}_1 \xi^{\alpha j}_1} + \cdots$. A Taylor expansion then leads to the form \eqn{QGPV} of $\barL Q$. 

It should not come as a surprise that the four mean-flow definitions lead to the same result: they all describe the same dynamics, specifically the same cluster of trajectories moving by $O(1)$ over a long time scale. The four Lagrangian-mean theories predict the position of this cluster using  notions of the centre of the cluster that differ by $O(\eps^2)$ only and indistinguishable to the order considered.

\section{Wave action} \label{sec:waveaction}

In the case where $\alpha$ is a phase-like variable ($\alpha \in \mathbb{S}^1$, where $\mathbb{S}^1$ is the circle) the flows $\phi^\alpha$ describe a closed loop in $\SDiff$. A conservation law is then associated with the invariance of the loop with respect to shifts in $\alpha$; this is the wave-action conservation law generalising the action conservation of classical mechanics \citep[e.g.][]{ArKoNe88}. We derive this conservation from the equations of motion, thus extending the derivation of \citet{AnMc78b} (see also \citet{Gr84} and \S10.3 of \citet{Bu14}).

Let $z^\alpha$ denote the divergence-free vector field generating shifts in $\alpha$ of the perturbation maps $\xi^\alpha$, that is,
\beq
\dt{}{\alpha} \,\xi^\alpha x = z^\alpha(\xi^\alpha x).
\eeq
\change{An application of \eqn{liecommute} shows} that
\beq \lab{dualpha}
\dt{}{\alpha} \, u^\alpha = \partial_t z^\alpha + \lie_{u^\alpha} z^\alpha . 
\eeq
Now, we pair the Euler equation \eqn{euler} for the flow $\nu = \nu^\alpha$, with $z^\alpha$ to obtain
\beq \lab{znu}
\change{\partial_t \nu^\alpha(z^\alpha) - \nu^\alpha(\partial_t z^\alpha) + \lie_{u^\alpha} \nu^\alpha (z^\alpha) - \nu^\alpha(\lie_{u^\alpha} z^\alpha) = - \lie_{z^\alpha} \pi^\alpha.}
\eeq
Using \eqn{dualpha}, we compute
\beq
\change{\nu^\alpha(\partial_t z^\alpha) = \nu^\alpha \left(\dt{}{\alpha}\,  u^\alpha - \lie_{ u^\alpha} z^\alpha\right) =  \dt{}{\alpha} \,  \tfrac{1}{2}\nu^\alpha(u^\alpha) - \nu^\alpha(\lie_{u^\alpha} z^\alpha).}
\eeq
Introducing this into \eqn{znu} leads to
\beq
\change{(\partial_t + \lie_{u^\alpha})\,  \nu^\alpha(z^\alpha) -  \dt{}{\alpha}\, \tfrac{1}{2} \nu^\alpha(u^\alpha) = - \lie_{z^\alpha}  \pi^\alpha.}
\eeq
Taking the average eliminates the derivative with respect to $\alpha$ and multiplying by the volume form $\omega$ gives 
\beq
\change{\partial_t \av{\nu^\alpha (z^\alpha)  } \omega
+ \av{  \lie_{u^\alpha}\nu^\alpha (z^\alpha)  \omega } 
+ \av{  \lie_{z^\alpha} \pi^\alpha \omega} = 0,}
\eeq
using also that $z^\alpha$ and $u^\alpha$ are divergence-free. Applying Cartan's formula \eqn{cartan} to write for example $ \lie_{z^\alpha} \pi^\alpha \omega = \d  (\pi^\alpha \ip_{z^\alpha} \omega)$, we finally obtain
\beq
\change{\partial_t \av{ \nu^\alpha(z^\alpha)  }\omega + \d \av{ \nu^\alpha(z^\alpha) \ip_{u^\alpha} \omega + \pi^\alpha \ip_{z^\alpha} \omega} = 0.}
\eeq
This is the local form of a conservation law for the action
\beq
A = \int_M \change{\av{ \nu^\alpha(z^\alpha) }} \, \omega , 
\eeq
since $u^\alpha \parallel \partial M$ and $z^\alpha \parallel \partial M$. Note that for small-amplitude perturbations $\xi^\alpha = \id + O(\eps)$, this action is $O(\eps^2)$ (since $\av{z^\alpha} = O(\eps^2)$) and is therefore a `wave activity' in the sense of \citet{AnMc78b}.

\section{Discussion} 

In this paper, we revisit \citeauthor{AnMc78a}'s GLM theory to remedy the deficiencies caused by its reliance on Euclidean parallel transport and consequent lack of  intrinsic  geometric meaning.  We achieve this by formulating the problem on a general Riemaniann manifold and by using the coordinate-free notation of modern differential geometry. Several lessons are learnt in the process. (i) The structure of the averaged equations that are obtained is  independent of the exact definition of the mean flow. (ii) These averaged equations are naturally expressed in terms of a Lagrangian-mean momentum $\barL \nu$, defined as the average of the pull-back of the momentum one-form (dual of the velocity field) from the perturbed to the mean configuration. It corresponds to
the quantity $\barL{\bu} - \textbf{\textsf{p}}$ of standard GLM, and is the integrand
of the mean Kelvin circulation theorem.
The central part played by the momentum one-form reflects its role as the natural dynamical variable of fluid dynamics.
(iii) The \change{mean flow is in fact defined as a mean flow map, from which the mean velocity is deduced by time differentiation. The definition of the mean flow map}  involves a degree of arbitrariness, as is inevitable when attempting to define the `mean' of an ensemble of points on a manifold (in this instance an ensemble of diffeomorphisms) that does not have a natural linear structure with natural averaging operators. Defining the mean  in terms of the flow map ensures that the mean flow retains its interpretation as a representative of an ensemble of nearby flows over long times.
(iv) It is clear that there are advantages, for incompressible fluids, in definitions that respect the volume-preservation constraint.  \change{For example in the case of MHD, the averaged equations \eqn{MHDniceave} retain the structure of the original MHD system.} (v) The pseudomomentum has a natural interpretation as a measure of the non-commutativity between two operations applied to the velocity field: Lagrangian averaging, and the transformation of vectors into one-forms by means of the metric. This measure encapsulates the impact of the perturbation on the mean flow. (vi) Our  computations of the mean flow and pseudomomentum show that it is possible to obtain explicit expressions perturbatively, order by order in the small-amplitude parameter $\eps$; the complexity of the resulting expressions suggests that in general they are probably best dealt with using a systematic method, for example a Lie-series expansion \citep[e.g.][]{Na73}. (vii) Among the intrinsic definitions of the mean that we consider, the glm theory of \citeauthor{SoRo10} appears to the most attractive for its relative ease of implementation. A potential weakness is however the fact that the vector field on which it relies is not \change{always} defined, even for near-identity flow, \change{though an approximate solution $q^\alpha$ of \eqn{glm} can computed order-by-order in $\eps$.} 
\change{The geodesic definition is likely more robust but leads to more complicated expressions. It is certainly the most natural of the other three definitions.}
(viii) In a perturbative context, the various definitions of the mean flow differ by bounded $O(\eps^2)$ terms, as expected since they all track (in somewhat different senses) the same ensemble of nearby flows. 

The aim of this paper is to introduce geometric generalised mean theories in as simple a set-up as possible. We envision extensions in several directions. A first direction concerns generalisations to fluid models more complex than the incompressible Euler equations on which the paper focuses. \change{Our brief treatment of stratified-rotating Boussinesq fluids,}  compressible fluids and MHD makes it plain that such generalisations are possible and likely advantageous. 
It would also be of interest to apply a geometric perspective to  commonly used averaging procedures other than GLM such as isentropic (or isopycnal) averaging. Another direction concerns manifolds with symmetries: key results of GLM and related theories apply to the special case where the ensemble average can be identified with an average along one coordinate (e.g., a zonal average), leading to the celebrated non-acceleration theorems, stemming 
from the mean Kelvin circulation theorem. The geometric understanding of these and their generalisation to a larger class of manifolds than  $\mathbb{R}^3$ and $\mathbb{S}^2$  will require us to study how continuous symmetries of manifolds can be exploited for averaging. A third direction concerns the use of geometric methods in the parameterisation of unresolved perturbations in numerical models. The geometric interpretation of the pseudomomentum, which captures this wave impact, can guide its parameterisation in terms of mean quantities as is required for the closure of the mean equations. In particular, it would be of interest to adapt the geometric set-up of the present paper to the short-wave, WKB treatment of the perturbations \citep[e.g.][]{GjHo96,BuMc98,BuMc05}.

\medskip
\noindent
\change{\textbf{Acknowledgements.} The authors are grateful to Andrew Soward for many useful discussions and for sending them papers in advance of formal publication, to Michael McIntyre for extensive comments and for encouraging the development of the Boussinesq example in \S\ref{sec:Bouss}, to  Darryl Holm for helpful comments and references, and to the other (anonymous) referee.  They thank Michael Singer for early conversations on differential geometry, Etienne Ghys and Oliver B\"uhler for helpful remarks, Marcel Oliver for comments that led to the restructuring of \S\ref{sec:smallamp}, and Steve Tobias for pointing out that they were working, at the time independently, along parallel tracks.}

\appendix

\section{Derivation details}

\noindent
\change{We set out geometric proofs of key identities involving Lie derivatives and time-dependent pull-backs; coordinate versions may also be found in \cite{SoRo10,SoRo14}.}

\subsection{Derivation of \eqn{dtpullback} and \eqn{pullbacklie}} \label{sec:pullback}

The first equality can be derived by \change{computing
\beq
\partial_t (\phi^*_{0,t} \tau) =\phi^*_{0,t} \left(\partial_t \tau + \partial_s|_{s=0} \left(\phi^*_{t,s} \tau \right) \right) = \phi^*_{0,t} \left(\partial_t \tau + \lie_u \tau \right),
\eeq
where, for clarity, we denote by $\phi_{t,s}=\phi_s \circ \phi_t^{-1}$ the map sending positions at time $t$ to positions at time $s$.}

The second equality, which expresses that the pull-back acts on Lie derivatives in a natural way can be derived as follows: denoting by $\psi_s$ the flow map associated with the velocity field $v$, we compute
\change{\beq
\phi^* \lie_v \tau = \phi^* \left.\dt{}{s}\right|_{s=0} \left(\psi_s^* \tau\right)
=  \left.\dt{}{s}\right|_{s=0} \left((\phi^{-1} \circ \psi_s \circ \phi)^* \phi^* \tau\right)
=  \lie_{\phi^* v} ( \phi^* \tau),
\eeq
where the last line relies on the definition of the Lie derivative and the observation that
\beq
\dt{}{s}\left( (\phi^{-1} \circ \psi_s \circ \phi) x\right) = (\phi^{-1})_* \left(v(\psi_s \phi x) \right) = (\phi^* v)(\psi_s x).
\eeq
}

\subsection{Derivation of \eqn{pullbackcons}} \label{sec:pullbackcons}

This property is readily established using \eqn{dtpullback}: applied to the flow $u^\alpha$ and the map $\phi^\alpha$ this equation implies that
\beq
(\partial_t  + \lie_{u^\alpha}) \tau^\alpha= 
\phi^{\alpha}_* \partial_t ( \phi^{\alpha*} \tau^\alpha) = 
\xi^{\alpha}_* \bar \phi_* \partial_t  ( \bar \phi^* \xi^{\alpha*} \tau^\alpha)
\eeq
on using \eqn{meanpert} and that $(\xi^\alpha \circ \bar \phi)^* = \bar \phi^* \xi^{\alpha*}$. Applying $\xi^{\alpha*}$ yields
\beq
\xi^{\alpha*} \left(\partial_t  + \lie_{u^\alpha}  \right)  \tau^\alpha = \bar \phi_* \partial_t ( \bar \phi^* \xi^{\alpha*} \tau^\alpha) = \left( \partial_t  + \lie_{\baru}  \right) (\xi^{\alpha*} \tau^\alpha),
\eeq
where the last equality follows again from \eqn{dtpullback}, this time applied to the tensor $\xi^{\alpha*} \tau^\alpha$ with the flow $\baru$ and map $\bar \phi$. An alternative derivation relying on \eqn{pullbacklie} is instructive. 
Using this equation, the left-hand side of \eqn{pullbackcons} can be written as
\begin{align}
\xi^{\alpha*} \left(\partial_t  + \lie_{u^\alpha}  \right)  \tau^\alpha &= \xi^{\alpha*} \partial_t \tau^\alpha  + \lie_{\xi^{\alpha*} u^\alpha} (\xi^{\alpha*} \tau^\alpha) \nonumber \\
 &= \partial_t (\xi^{\alpha*} \tau^\alpha) - \xi^{\alpha*} \lie_w  \tau^\alpha + \lie_{\xi^{\alpha*} u^\alpha} (\xi^{\alpha*} \tau^\alpha) \nonumber \\
&= \partial_t (\xi^{\alpha*} \tau^\alpha)  - \lie_{\xi^{\alpha*} w} (\xi^{\alpha*} \tau^\alpha) + \lie_{\xi^{\alpha*} u^\alpha} (\xi^{\alpha*} \tau^\alpha)  \nonumber \\
&= \partial_t (\xi^{\alpha*} \tau^\alpha) + \lie_{\xi^{\alpha*} (u^\alpha-w)} (\xi^{\alpha*} \tau^\alpha) = \left( \partial_t  + \lie_{\baru}  \right) (\xi^{\alpha*} \tau^\alpha).
\end{align}
Here we have used \eqn{dtpullback}, \eqn{pullbacklie} and \eqn{uuw} to write the second line, third and fourth line, respectively.

\section{Euler equations as geodesics} \label{sec:varieuler}

The Euler equations in the form \eqn{euler}  are obtained as geodesic equations on $\SDiff$ considered as an (infinite-dimensional) Riemannian manifold when equipped with the metric associated with the distance \eqn{D22}. To show this, we consider the augmented action
\beq \label{augaction}
\mathscr{A}[\phi,\pi] =  \int_0^T  \int_\M \left( \tfrac{1}{2} g(u,u) \omega + \pi \left( \phi^* \omega - \omega \right)\right) \, \d t,
\eeq
where $\pi$  is a scalar field with the role of a Lagrange multiplier enforcing volume preservation. This action is required to be stationary for perturbations $\phi \mapsto \psi_s \circ \phi $ with arbitrary, $t$-dependent $\psi_s$ such that 
$\psi_s|_{t=0} = \psi_s|_{t=T}=\id$.
This perturbation changes the velocity field $u$ into
\change{
\beq
u_s(x)= \dt{}{t} (\psi_s \circ \phi) \circ \phi^{-1} \circ \psi_s^{-1}.
\eeq
Denoting by $v$ the vector field generating $\psi_s$ as a function of $s$, in the sense that
\beq \lab{varv}
\left.\dt{}{s}\right|_{s=0} \psi_s x = v(x),
\eeq
we can apply \eqn{liecommute} to the two-parameter map $\psi_s \circ \phi$ and obtain the first-order variation of $u_s$ as
\beq \lab{B4}
\left.\dt{}{s}\right|_{s=0} u_s = \partial_t v - \lie_v u = \partial_t v + \lie_u v.
\eeq}
Note that we require $v \parallel \partial M$, as for $u$,  but $v$ need not be divergence free.

With this result, we  compute
\beq \lab{var}
\left.\dt{}{s}\right|_{s=0} \mathscr{A}[\psi_s \circ \phi,\pi_s] 
= \int_0^T \!\! \int_\M \left(\nu (\partial_t v + \lie_u v) \omega + \pi \lie_v \omega 
\right) \, \d t.
\eeq
This is rewritten using integration by parts, and the volume preservation $\phi^* \omega = \omega$ enforced by the Lagrange multiplier $\pi$ and its consequence $\lie_u \omega = 0$.  
First, we have that
\beq \lab{intbp1}
\int_0^T \int_M \nu (\partial_t v)  \omega \, \d t= - \int_0^T \int_M \change{\partial_t \nu(v)} \,\omega \, \d t
\eeq
since $v|_{t=0}=v|_{t=T}=0$. Second, 
\beq \lab{intbp2}
\int_M \nu (\lie_u v ) \omega = \int_M \lie_u ( \change{\nu(v)} \, \omega) - \int_M \change{(\lie_u \nu)(v)} \, \omega = \int_{\partial M} \change{\nu(v)} \ip_u \omega  -  \int_M \change{(\lie_u \nu)(v)} \, \omega, 
\eeq 
where we have used Cartan's formula \eqn{cartan} to write $\change{\lie_u ( \nu(v) \omega) = \d (\nu(v)} \ip_u \omega)$ and Stokes' theorem \citep[e.g.][]{Sc80,Fr97}.  The boundary integral vanishes since $u \parallel \partial M$. Third, we have similarly
\beq \lab{intbp3}
\int_M \pi \lie_v \omega = \int_M \lie_v (\pi \omega) - \int_M  ( \lie_v \pi )\omega = \int_{\partial M} \pi  \ip_v \omega - \int_M \change{\d \pi(v)} \, \omega,
\eeq
with again a vanishing boundary integral, this time because $v \parallel \partial M$. 
Introducing \eqn{intbp1}--\eqn{intbp3} into \eqn{var} gives
\beq
\left.\dt{}{s}\right|_{\epsilon=0} \mathscr{A}[\psi_s \circ \phi,\pi_s] 
= - \int \!\! \int_\M  \change{\left(\partial_t \nu + \lie_u \nu + \d \pi \right)(v)} \,
\omega \, \d t.
\eeq
This variation vanishes for arbitrary $v$  leading to the Euler equations \eqn{euler}, with the condition $\lie_u \omega = 0$ written as $\divv u =0$ by definition of the divergence.

\change{Note that the derivation in this appendix, in particular the constrained variations \eqn{B4}, can be  generalised to cover the class of so-called Euler--Poincar\'e equations that includes, among others, all the  fluid models treated in this paper; see \cite{HoMaRa98} for details.}

\section{Variational derivation of \eqn{aveuler}} \label{sec:GLMlagr}

Substituting the decomposition \change{\eqn{uuw}} into the action \eqn{action}, enforcing the volume-preservation constraints for $\xi^\alpha$ and $\bar \phi$ by means of Lagrange multipliers $\pi^\alpha$ and $\bar \pi$ and averaging leads to
\beq  \lab{avaction}
\av{\mathscr{A}}[\xi^\alpha,\bar \phi, \pi^\alpha,\bar \pi] =  \av{ \int_0^T  \int_\M \left( \tfrac{1}{2} g(w^\alpha+\xi_*^\alpha \baru,w^\alpha+\xi_*^\alpha \baru) \omega + \pi^\alpha \left( \xi^{\alpha*} \omega - \omega \right)
+ \bar \pi \left( \bar \phi \omega - \omega \right)
\right) \, \d t}.
\eeq
Requiring this to vanish for arbitrary variations of $\xi^\alpha$,  $\bar \phi$, $\pi^\alpha$ and $\bar \pi$, yield dynamical equations for both the mean and perturbations. First, variations with respect to $\bar \phi$ generated by a vector field $v$ as in \eqn{varv} lead to
\beq \lab{varapp}
\av{ \int_0^T  \int_\M \change{\nu^\alpha \left(\xi_*^\alpha (\partial_t v + \lie_{\baru} v)\right)} \, \omega +  \bar \pi \lie_{v} \omega}=0. 
\eeq
Now, we can pull-back the first term to write
\beq
\change{\av{\int_\M  \ \nu^\alpha \left( \xi^{\alpha}_* (\partial_t v + \lie_{\baru} v) \right) \, \omega }
= \av{\int_\M  (\xi^{\alpha *} \nu^\alpha) \left(\partial_t v + \lie_{\baru} v \right) \, \omega }
= \int_\M  \barL{\nu} \left(\partial_t v + \lie_{\baru} v \right) \, \omega.}
\eeq
Introducing this into \eqn{varapp} and integrating by parts in the manner of appendix \ref{sec:varieuler} gives the averaged equation \eqn{aveuler}, with $\bar \pi$ replacing $\barL{\pi}$. 
Variations of \eqn{avaction} can be shown to lead to the Euler equations \eqn{euler} for each of the momenta $\nu^\alpha$.

\section{Riemannian centre of mass} \label{app:frechet} 

Consider a manifold $\M$ with a distance between points $x$ and $y$ given by $d(x,y)$. Suppose points $x^\alpha$ are distributed according to a measure $\mu(x) \, \d x$. Then the Riemannian centre of mass (Fr\'echet or Karcher mean) of their position is defined as
\beq
\bar x  = \argmin_{ x \in M} \av{d^2( x,x^\alpha)} =  \argmin_{ x \in M} \int_M d^2( x,y) \, \mu(y) \, \d y. \lab{frechet}
\eeq
The existence and uniqueness of such a minimiser is of course not trivial and depends on both the manifold $\M$ and  measure $\mu(x) \, \d x$.

One interpretation of the centre of mass is that it provides the best approximation of the measure $\mu(x)\, \d x$ by an atomic measure $\bar{\mu}(x) \,\d x = \delta(x-\bar x) \,\d x$ when the Wasserstein distance $W^2(\mu,\nu)$ between the measures is used to define the approximation. This distance is defined as
\[
W^2(\bar \mu,\mu) = \inf_{k(x,y)} \int d^2(x,y) k(x,y) \, \d x \,\d y,
\]
with $\int k(x,y) \, \d y = \bar \mu(x) = \delta( x- \bar x )$ and $\int k(x,y) \, \d x = \mu(y)$. It is clear that $k(x,y)= \delta(\bar x - x) \mu(y)$, so that the minimisation reduces to
\[
W^2(\bar \mu,\mu) = \inf_{x \in M} \int_M d^2( x,y)\, \mu(y) \, \d y,
\]
in agreement with the definition \eqn{frechet} of the centre of mass.

The connection between the optimal transport and geodesic definitions of \change{the mean flow map} of \S\S\,\ref{sec:OT}--\ref{sec:geodesic} is made clear by thinking of the ensemble of flow maps $\phi^\alpha$ as defining a measure on the space of diffeomorphisms $\SDiff$; for example, for a discrete family labelled by $\alpha=1,2,\ldots$, this is the atomic measure concentrated on the maps $\phi^\alpha \in \SDiff$. Such a measure on $\SDiff$ is in fact a more general concept since it allows for continuous, multidimensional $\alpha$, for instance, and it arises in the theory of generalised flows \citep{Br89,Sh94}. From this point of view the problem of defining \change{the mean flow map $\bar \phi$} corresponds to specifying a classical flow that is `closest' to a given generalised flow.

\section{Derivation of \eqn{pressureless}, \eqn{pmean} and \eqn{pmeanOT}} \label{sec:variother}

The extended GLM formulation  \eqn{GLM} requires that the action
\beq \lab{GLMaction}
\mathscr{A}[\gamma_s] = \av{ \int_0^1  \int_\M \left( g(q^\alpha,q^\alpha) \gamma^{\alpha}_{s*} \omega \right) \, \d s \, },
\eeq
be stationary for perturbations $\gamma^\alpha_s \mapsto \psi^\alpha_{\eps} \circ \gamma^\alpha_s$ of the paths joining $\bar \phi$ to $\phi^\alpha$. Here the perturbation paths $\psi^\alpha_\eps$ are parameterised (implicitly) by $s$ 
and  constrained by the condition $\psi^\alpha_\eps|_{s=1}=\id$ at the endpoint $s=1$. At the other endpoint $s=0$, the constraint is weaker since that starting map is not fixed but an arbitrary perturbation of $\bar \phi$; thus 
$\psi_\eps^\alpha|_{s=0}$
 is only required to be independent of $\alpha$.  Denoting by $v^\alpha$ the vector field generating $\psi^\alpha_\eps$ in a manner analogous to \eqn{varv}, these constraints become 
$v^\alpha|_{s=1}=0$ and $v^\alpha|_{s=0}=v$, 
where $v$ is an arbitrary vector field independent of $\alpha$. 
The velocity fields $q^\alpha$ appearing in \eqn{GLMaction} are perturbed to $q^\alpha_\eps$; \change{using \eqn{liecommute}, we have then that}
\beq
\left.\dt{}{\epsilon}\right|_{\epsilon=0} q^\alpha_\eps = \partial_s v^\alpha + \lie_{q^\alpha} v^\alpha.
\eeq
Using this, we can compute the first-order variation $\d \mathscr{A} / \d \eps |_{\eps=0}$ and require that it vanishes. The computation is very similar to that in appendix \ref{sec:varieuler} and we do not detail it here. It leads to the pressureless equation in the form \eqn{euler20}, and to the boundary term 
\beq \lab{bound}
\change{\av { \qflat^\alpha (v^\alpha) } = \av{\qflat^\alpha} (v)} \ \ \textrm{at} \ \ s= 0.
\eeq
Since $v$ is arbitrary, this gives constraint \eqn{pmean} on the initial value of $\qflat^\alpha$. 

The optimal transportation formulation differs from the extended GLM one only by the addition of the constraint that $\bar \phi$ be volume preserving. To enforce this, we augment the action \eqn{GLMaction} by the term
\beq
- \,\av { \int_M \psi (\bar \phi^* \omega - \omega)\, },  
\eeq
where $\psi$ is a Lagrange multiplier. Using that 
$\bar \phi \mapsto \psi^\alpha|_{s=0} \circ \bar \phi$,
 we find that this extra term combines with the boundary term \eqn{bound} to give
\beq
\change{ \av { \qflat^\alpha (v^\alpha) - \d \psi (v^\alpha) } = \av{\qflat^\alpha - \d \psi} (v)  }\ \ \textrm{at} \ \ s= 0, 
\eeq
leading to condition \eqn{pmeanOT}.

\bibliographystyle{jfm}

\end{document}